\documentclass[proof]{WileyASNA-v1}

\articletype{ORIGINAL ARTICLE}%

\received{}
\revised{}
\accepted{}

\begin{document}

\title{The ORCA-TWIN qCMOS Experiment I. \\ Science case and commissioning at Calar Alto Observatory}

\author[1,2,3]{Martin M. Roth*}
\author[2]{Paško Roje}
\author[2]{Stella Vješnica}
\author[5,7]{Stefan Cikota}
\author[4]{Alex J. Brown}
\author[2,6]{Mike Kretlow}
\author[5]{Marco Azzaro}
\author[5]{Santiago Reinhart}
\author[5,6]{Jes\'us Aceituno}
\author[3]{Harry Dawson}
\author[1,3]{Carsten Denker}
\author[2,9]{Torsten Enßlin}
\author[3]{Stephan Geier}
\author[1]{Thomas Granzer}
\author[4,8]{Thomas Kupfer}
\author[11]{Gianluca Li Causi}
\author[2,3,10]{Samaya Nissanke}
\author[6]{José Luis Ortiz}
\author[11,12]{Fernando Pedichini}
\author[1,3]{Katja Poppenhäger}
\author[1,3]{Axel Schwope}

\authormark{Roth \textsc{et al}}

\address[1]{\orgname{Leibniz-Institut für Astrophysik Potsdam (AIP)}, \orgaddress{\state{An der Sternwarte 16, 14482 Potsdam}, \country{Germany}}}

\address[2]{\orgname{Deutsches Zentrum für Astrophysik (DZA)}, \orgaddress{\state{Postplatz 1, 02826 Görlitz}, \country{Germany}}}

\address[3]{\orgdiv{Institut für Physik und Astronomie}, \orgname{Universität Potsdam}, \orgaddress{\state{Karl-Liebknecht-Str. 24/25, 14476 Potsdam}, \country{Germany}}}

\address[4]{\orgdiv{Hamburger Sternwarte}, \orgaddress{\state{Gojenbergsweg 112, 21029 Hamburg}, \country{Germany}}}

\address[5]{\orgdiv{Observatorio de Calar Alto},  \orgaddress{\state{Sierra de los Filabres, E-04550 Gérgal}, \country{Spain}}}

\address[6]{\orgname{Instituto de Astrof\'{i}sica de Andaluc\'{i}a (IAA-CSIC)}, \orgaddress{\state{Glorieta de la Astronom\'{i}a, s/n, 18008 Granada}, \country{Spain}}}

\address[7]{\orgdiv{Faculty of Electrical Engineering and Computing, Department of Communication and Space Technologies}, \orgname{University of Zagreb}, \orgaddress{\state{Unska 3, 10000 Zagreb}, \country{Croatia}}}

\address[8]{\orgdiv{Department of Physics and Astronomy}, \orgname{Texas Tech University}, \orgaddress{\state{Lubbock, TX 79409-1051}, \country{USA}}}

\address[9]{\orgname{Max Planck Institute for Astrophysics}, \orgaddress{\state{Karl-Schwarzschildstr. 1, 85741 Garching}, \country{Germany}}}

\address[10]{\orgname{Deutsches Elektronen Synchrotron DESY}, \orgaddress{\state{Platanenallee 6, 15738 Zeuthen}, \country{Germany}}}

\address[11]{\orgname{$^{11}$Osservatorio Astronomico di Roma}, \orgaddress{\state{00078 Monte Porzio Catone (RM)}, \country{Italy}}}

\address[12]{\orgdiv{ADONI, ADaptive Optics National laboratory of Italy}, \orgname{INAF}, \orgaddress{\state{} \country{Italy}}}

\corres{*Martin M. Roth, \email{mmroth@aip.de}}

%\presentaddress{This is sample for present address text this is sample for present address text}

\abstract{
We describe a pilot study to explore a new generation of fast and low noise CMOS image sensors for time domain astronomy, using two remote telescopes with a baseline of 1635~km. The
experiment involves direct imaging with novel qCMOS image sensor technology that combines fast readout with sub-electron readout noise. Moreover, synchronized observations from two remote telescope sites will be used to explore new approaches for measuring Solar System bodies, precision stellar photometry, and speckle imaging. A fast-track installation of an ORCA-Quest\,2 camera at the Calar Alto Observatory 1.23m telescope has demonstrated the potential of the qCMOS technology for time domain astronomy. Numerical simulations suggest that owing to sub-electron readout noise, qCMOS sensors outperform classical CCDs for high-cadence imaging on 1m-class telescopes. The small penalty for post-readout binning, that is almost insignificant in comparison to higher readout noise detectors, opens interesting applications for scene-dependent data processing in direct imaging, and potentially even for spectroscopy.}


\maketitle

\clearpage

\section{Introduction} \label{sec:introduction}

From a historical perspective, charge coupled devices (CCD) have enabled from the 1980s a breakthrough for astronomy in the visible, with impact on almost all kinds of observing techniques, from direct imaging, over spectroscopy, to high precision polarimetry, whether in space, or at ground-based observatories. After three decades of development and constant improvement, the modern CCD must be considered a mature technology, and an almost ideal detector for scientific applications, providing high quantum efficiency (QE), low read noise, excellent linearity, and a large dynamic range \citep{Roth2023}. The one drawback for time-domain astronomy, however, is its principle of operation: charge transfer means a slow readout process through the bottleneck of one (or a few) output nodes. For fast cadence observations, the deadtime for readout between exposures makes it prohibitive to use full frame CCDs for cadence shorter than 10~s.

As a variant of the classical CCD that overcomes this disadvantage, the Electron Multiplication CCD (EMCCD, also LLLCCD for low light level CCD, or L3CCD) was introduced by \citet{Jerram2001} as a sensor where individual photoelectrons are amplified in a series of high voltage electrodes, resulting in ionization of secondary electrons and a cascade of charge with an output node voltage that is much higher than the read noise, thus the ability to detect single photons. Therefore, EMCCDs have found use for astronomical applications with very short exposure times and high cadence. The properties of EMCCDs in comparison with CCDs are discussed in detail by \citet{Smoith2004,Mackay2004,Burke2005}. In summary, the EMCCD features sub–electron readout noise, high readout speed, and high quantum efficiency. Amongst the drawbacks, it exhibits multiplication noise, a low dynamic range, and an undesirable background from clock induced charge.

From the early 2000s, CMOS image sensors were beginning to enter the arena of scientific detectors, e.g.\ \citet{Janesick2002,Ay2002}. A comprehensive summary of CMOS properties was presented in a series of conference papers by \citet{Janesick2006,Janesick2007,Janesick2009,Janesick2010}. As opposed to the slow serial readout of the CCD, the architecture of this type of sensor (also alluded to as ``active pixel sensor'') provides a readout node in each pixel. Multiplex access to the individual pixel charges/voltages allows for fast readout of the entire sensor, and frame rates of order 100~Hz, or even faster. Detectors that are, as opposed to consumer applications, optimized for scientific imaging, are often called sCMOS. The use of sCMOS sensors for space applications was discussed by \citet{Jerram2020}. A critical review of sCMOS sensors in comparison to CCD for astronomy was presented by \citet{Hodapp2006}. At present, this type of detector is predominantly found in space applications, e.g. remote sensing, but on the ground it is still largely limited to amateur astronomy.

An interesting sCMOS sensor development from BAE Systems with high QE (70\% at 600~nm), low read noise (1~e$^-$), and fast readout (100~Hz) was presented by \citet{Vu2012}. It has been further optimized and is now available as a $4096\times2304$ format, 4.6~$\mu$m  pixel back-side illuminated device (BAE HWK 4123) from Fairchild Imaging (the company was renamed in 2024 after acquisition by Hamamatsu). The astronomical image sensor community was made aware of an off-the-shelf camera system available from Hamamatsu, based on this chip, at the Scientific Detector Workshop 2022, which had a peculiar focus on quanta sensors that can detect single photons, such as skipper CCDs, or low-noise CMOS \citep{Amico2023}. Although two years later applications for speckle interferometry at the Sternberg 2.5m telescope \citep{Strakhov2024} and for the Visible Aperture-Masking Polarimetric Imager/Interferometer with extreme adaptive optics at the Subaru telescope \citep{Lucas2024} were reported to have made use of the Hamamatsu camera, that comes with the brand name ORCA-Quest,\footnote{\url{https://www.hamamatsu.com/us/en/product/cameras/qcmos-cameras/C15550-20UP.html}} not many other applications for astronomy are known to date. 

The technology has found interest for space applications \citep{Krynski2025b}, also with regard to another device, the GigaJot GJ01611, that used to be commercially available in the GigaJot QIS16 camera \citep{Ma2021}, which, however, has disappeared from the market. This back-illuminated sensor featured very low read-noise of 0.19~e$^-$ in a format of $4096\times4096$ pixels, however at the expense of tiny pixels with a pitch of 1.1~$\mu$m  \citep{Krynski2025a}.

Because of the relevance for future developments, the term qCMOS will be used in the following as a general definition for \emph{very low read-noise, quanta CMOS sensors}, regardless of commercial branding. The ORCA-TWIN experiment makes use of the model ORCA-Quest\,2 camera that employs the BAE HWK 4123 qCMOS sensor mentioned above.

The German Center for Astrophysics (DZA) in Görlitz, Germany, whose mission is to perform, firstly, fundamental research in astrophysics, secondly, technology development for astronomical instrumentation, and thirdly, big data analysis, discovered that the commercially available ORCA-Quest camera would motivate interesting science cases for time domain astronomy within the envisaged research portfolio, and possibly a fast track pilot study at 1m-class telescopes. Here, we report on the implementation of this idea within the ORCA-TWIN experiment, and the first steps to realization.

The paper is organized as follows: Section~\ref{sec:science} introduces the science case and concept, Sect.~\ref{sec:qcmos} discusses the advance of the qCMOS technology over classical image sensors, Sect.~\ref{sec:simulations} provides numerical simulations of performance at the telescope, Sect.~\ref{sec:commissioning} describes the commissioning of one of two cameras at Calar Alto, and Sect.~\ref{sec:results} summarizes the results from the first observations. We close the paper with conclusions and an outlook in Sect.~\ref{sec:conclusions}. 

%--------------------------------------------------------------------

\section{ORCA-TWIN experiment} 
\label{sec:science}

Motivated by an initial focus of the DZA research agenda on multi-messenger and time domain astronomy, also the vision to engage the regional semiconductor eco-system around Dresden for the development of novel image sensors, the idea was spawned to test the novel fast, low noise Hamamatsu ORCA-Quest\,2 technology for high cadence synchronized observations of the same object with two telescopes at remote sites. This has led to the definition of the ORCA-TWIN pilot project, involving the 1.23m telescope at Calar Alto Observatory (CAHA), and one of the pair of 1.2m robotic STELLA telescopes at the Teide Observatory, Tenerife. 

The hypotheses that shall be tested are, firstly, the ability to directly triangulate near Earth Solar System objects, and, secondly, the expected capability to reduce the effects of atmospheric turbulence and extinction through simultaneous observations from two sites.

Given the swift interest from the user community, the ORCA-TWIN experiment was formally granted as a DZA pilot project in March 2025. In collaboration with Calar Alto Observatory and the STELLA telescopes on Teide Observatory, operated by Leibniz Institute for Astrophysics Potsdam, it was decided to implement the project on a fast track. A report on commissioning at Calar Alto in June 2025 is presented in Section~\ref{sec:commissioning}. A report on results from ongoing laboratory test will be presented in Paper~II. The installation of the second camera at STELLA is planned for early 2026, to be presented in Paper~III, including simultaneous operation of the two telescopes. In what follows, we briefly outline science cases that were put forward to justify the proposal for the pilot project. Note that some of these science cases strictly require simultaneous observations from two sites, whereas others do not.

\subsection{Triangulation of near-Earth objects}
\label{subsec:triangulation}

\begin{figure}
	\centering
   \includegraphics[width=0.48\columnwidth]{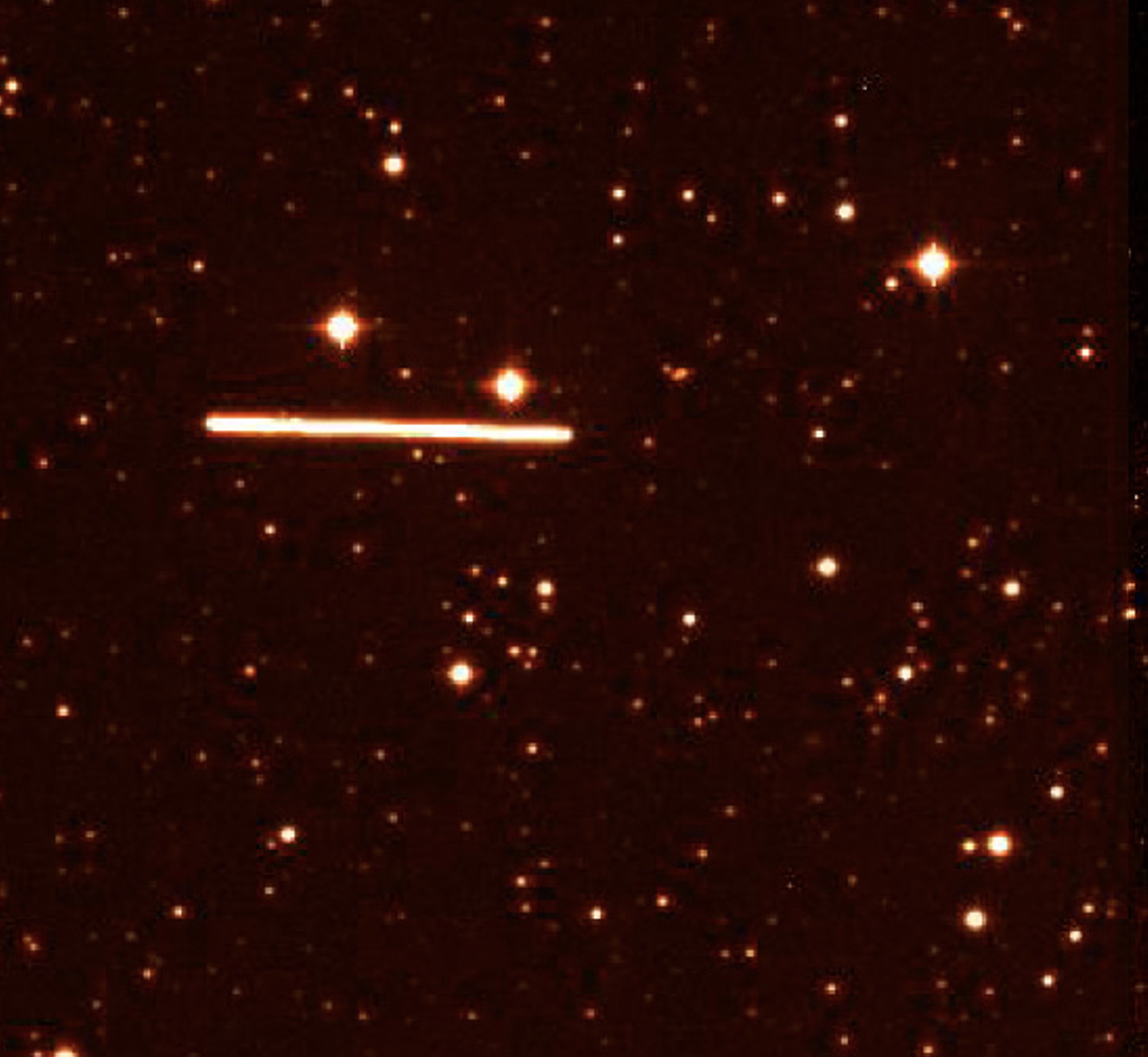}
    \hspace{1mm}
    \includegraphics[width=0.48\columnwidth]{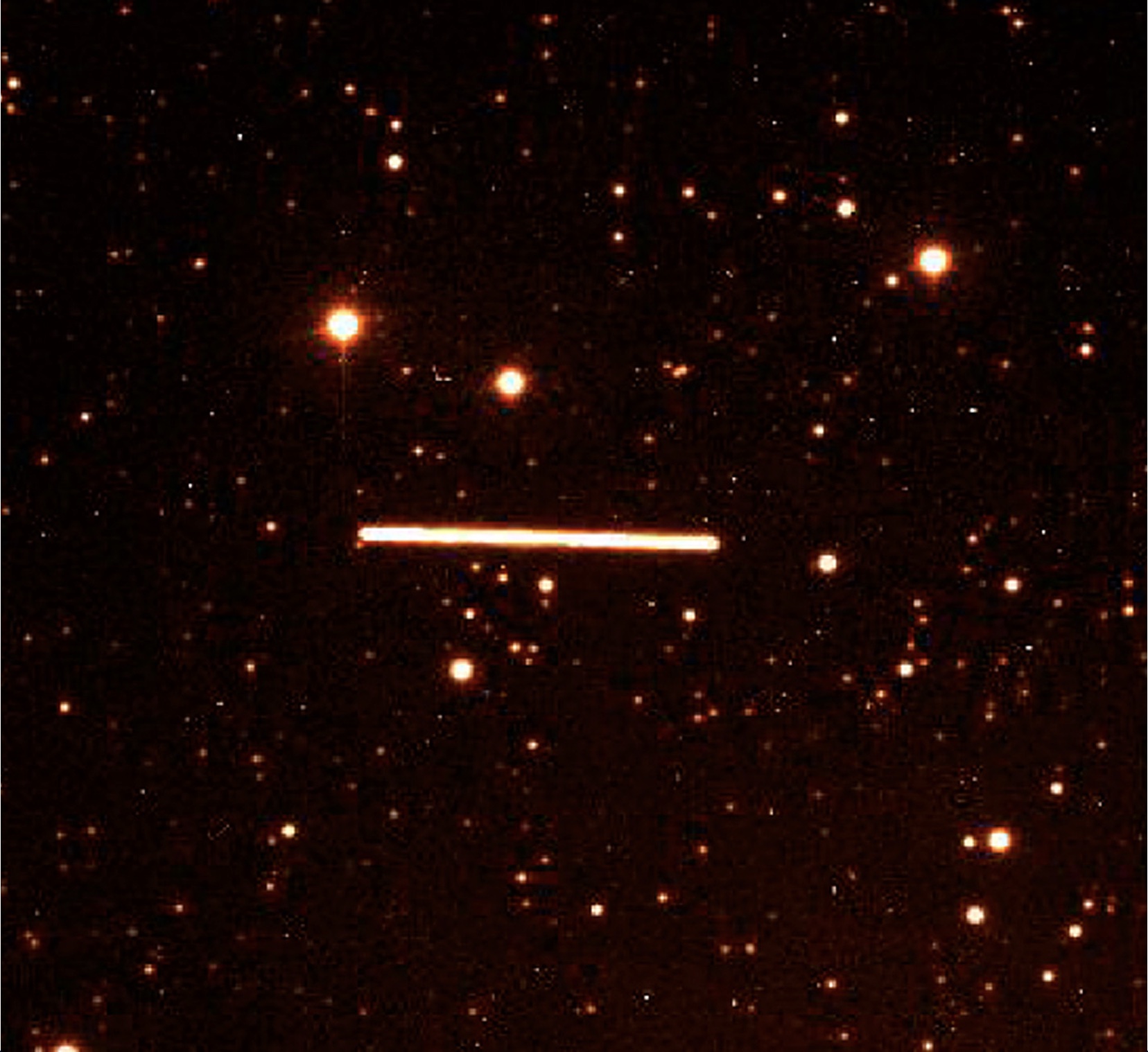}
	\caption{Simultaneous observations of asteroid 4179 Toutatis, left: WFI La Silla, right: FORS1 Paranal. Credit: ESO}
	\label{fig:toutatis}
\end{figure}

\begin{figure}
	\centering
	\includegraphics[width=\columnwidth]{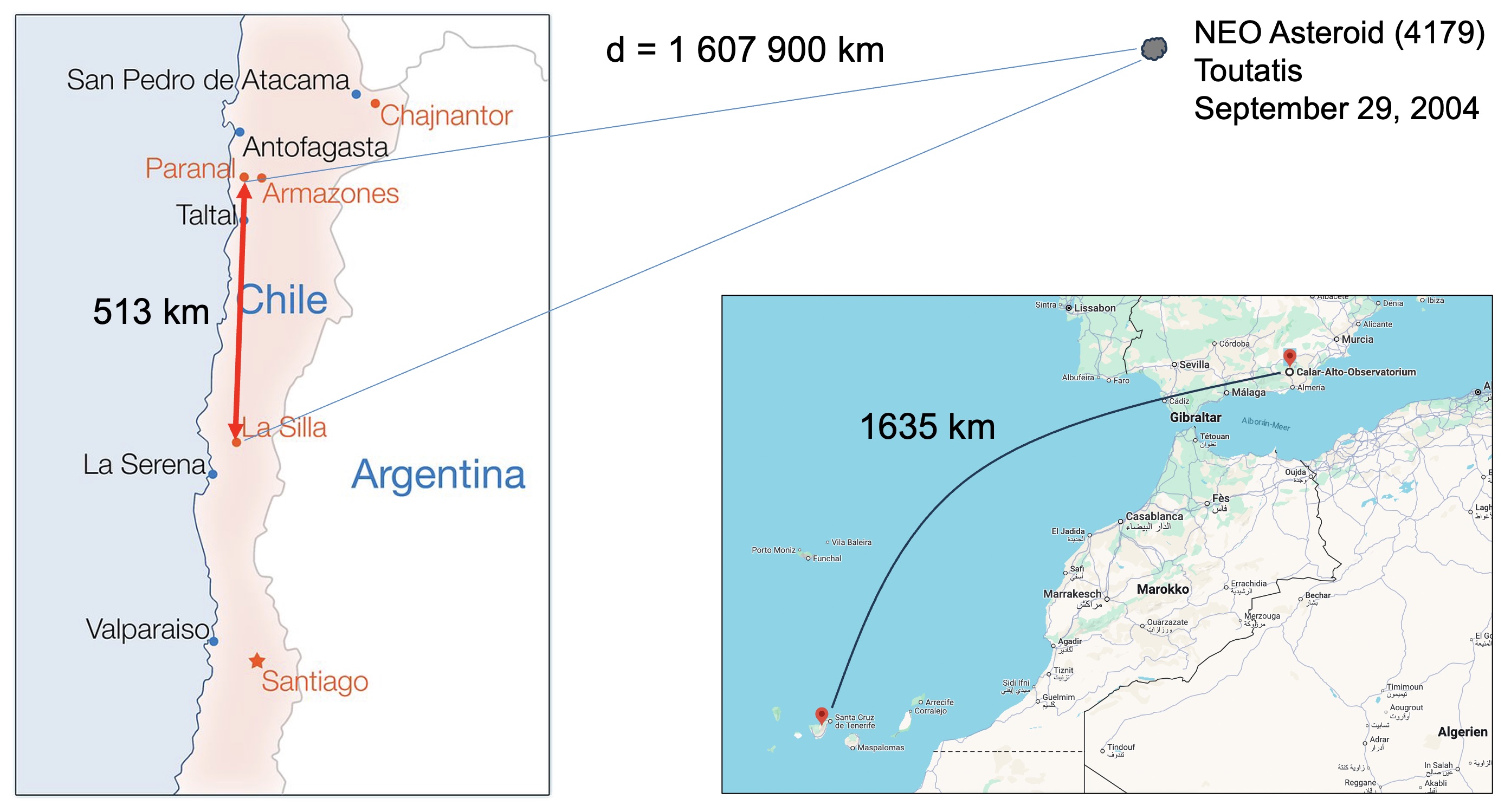}
	\caption{Left: triangulation of asteroid 4179 Toutatis with 513~km baseline between ESO observatories at Paranal and La Silla. Right: 1635~km baseline of ORCA-TWIN sites on mainland Spain and Tenerife. Credit: ESO, OpenStreetMap Wiki.}
	\label{fig:baseline}
\end{figure}

The possibility to directly measure the distance of near-Earth objects (NEOs), i.e.\ Solar System bodies with a perihelion of less than 1.3 astronomical units (AU), was demonstrated with simultaneous observations of the near-Earth passage of asteroid 4179 Toutatis from ESO observatories Paranal and La Silla on September 29, 2004, at 02:30 hrs UTC, when the object was predicted to miss the earth by merely 1\,640\,000~km, i.e.\ four times the distance of the Moon.\footnote{ESO press release ESO0403 on September 29, 2004} Figure~\ref{fig:toutatis} shows the tracks of the fast moving object, with an apparent magnitude of roughly 10~mag, and a velocity of 11 km\,s$^{-1}$ relative to the earth, obtained from 1-minute exposures through an R filter with the Wide Field Imager (WFI) at the ESO/MPG 2.2m telescope on La Silla (left), and the same through a narrow-band [O\,III] filter with the FORS1 instrument at UT2 on Paranal (right). From the parallax of about 40\,arcsec\ and a distance of the observatories of 513~km, the distance to the asteroid could be readily computed as 1\,607\,900~km (Figure~\ref{fig:baseline}). 

We assume that under similar circumstances, but considerations of a more than three times longer baseline, and sub-arcsecond precision astrometry that will become possible with short dual ORCA-TWIN exposures, synchronized with a GPS standard to millisecond accuracy, the range of triangulation can be significantly larger than the Toutatis example, which would be useful for newly detected NEOs for which Radar ranging measurements are not available. The mean parallax for our baseline and a distance $\Delta$ of 1\,AU is 1.8018\,arcsec. With an assumed astro\-metric precision of 100\,mas, or 50\,mas, respectively, we estimate triangulation distance uncertainties as listed in Table~\ref{tab:disterror}.

\begin{table}[h!]
\centering
\caption{Relative distance uncertainty}
\begin{tabular}{ccc}
\hline
$\Delta$ (AU) & $\sigma\Delta$/$\Delta$ (50 mas) [\%]	&  $\sigma\Delta$/$\Delta$ (100 mas) [\%]	\\
\hline
0.1	   &   0.39             & 	0.78                \\
0.5    &   2.0              &   3.9                 \\
1.0	   &   3.9              &   7.8                 \\
3.0	   &   11.8             &	23.6                \\
5.0	   &   19.6             &   39.2                \\
\hline
\end{tabular}
\label{tab:disterror}
\end{table}

With similar arguments, low Earth orbit objects (LEO), satellites, and also space debris would become accessible to fast cadence ORCA-TWIN observations as opposed to streak detection \citep{Tonry2011,Nir2018,Wang2022}.

\subsection {Photometry of very fast rotating near-Earth objects} 
\label{subsec:fastrotNEO}

\begin{figure}
	\centering
	\includegraphics[width=\columnwidth]{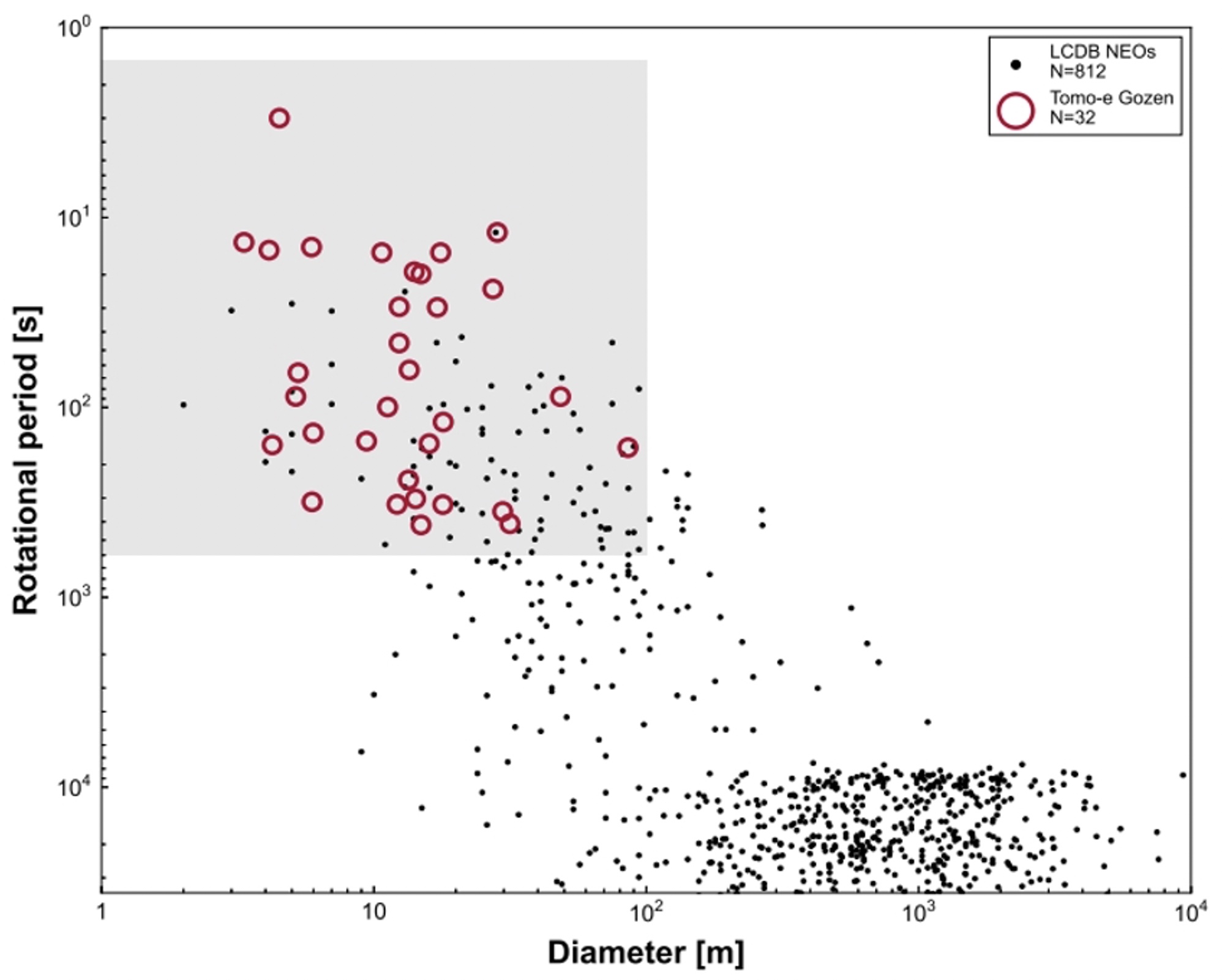}
	\caption{The D--P diagram (diameter vs.\ rotational period) shows the so-called 2.2-hour spin barrier for asteroids larger than about 150\,--\,200~m, but also displays a possible cut-off at $P \approx 10$~s (with one exception). Figure adopted from \citet{Beniyama2022}.}
	\label{fig:rot-NEO}
\end{figure}

Most asteroids have rotation periods ranging from several hours to a few days. However, a subset of small NEOs exhibits ultra-fast rotation, with periods of just minutes or even seconds. Both ends of the rotational period spectrum -- extremely slow rotators (with periods on the order of weeks) and ultra-fast rotators -- are subject to strong observational biases. In particular, ultra-fast rotators are underrepresented in existing datasets, largely because traditional asteroid surveys typically operate with cadences of several minutes or more, rendering them insensitive to sub-minute variations. To reliably detect and characterize such fast-spinning bodies, high-cadence time-series photometry with sufficient signal-to-noise ratio (SNR) is essential. NEOs offer a unique opportunity to investigate the physical and structural properties of small bodies, especially when they make close approaches to Earth and become accessible to small- and medium-aperture telescopes such as the CAHA 1.23m telescope. The rotation period of a small asteroid encodes information about its dynamical evolution and internal strength, as small bodies are particularly susceptible to rotational acceleration via the YORP (Yarkovsky--O'Keefe--Radzievskii--Paddack) effect. There exists a well-known spin barrier at about 2.2\,h, beyond which asteroids larger than about 150\,--\,200\,m in diameter rarely rotate faster (Fig.~\ref{fig:rot-NEO}). This threshold reflects the cohesionless structural limit of so-called ``rubble-pile'' asteroids -- aggregates of collisional debris held together by gravity. Asteroids smaller than this threshold are likely monolithic fragments capable of sustaining much faster rotations due to internal material strength. If such a rubble-pile object were spun up beyond this critical limit, it would likely undergo structural failure, potentially leading to mass shedding or binary formation \citep{Pravec2007}.

While the population of asteroids with rotation periods above the spin barrier is well-sampled and continues to grow, the population of fast rotators remains sparse and statistically incomplete. Expanding the sample of ultra-fast rotators is therefore essential. Notably, \citet{Beniyama2022} reported a sharp truncation in the diameter-period (D--P) distribution of small NEOs at rotation periods around 10~s (Figure~\ref{fig:rot-NEO}). This cutoff, which appears as a flat-topped distribution, cannot be easily explained by existing models based on tensile strength or YORP suppression via meteoroid impacts, suggesting the need for additional observational data to test and refine these hypotheses. We propose an observing program targeting very small NEOs (absolute magnitude H $>$ 22, corresponding to diameters $\le200$~m), using high-cadence photometry tailored to each object's brightness. Depending on the apparent magnitude, cadence rates from about 0.1~Hz up to 2\,--\,4~Hz will be employed to identify and characterize rotation periods in the range of several minutes down to under 10~s.

\subsection{Measuring sub-milliarcsecond stellar angular sizes from diffraction patterns during asteroid occultations}
\label{subsec:occult-stars}

\begin{figure}
	\centering
	\includegraphics[width=\columnwidth]{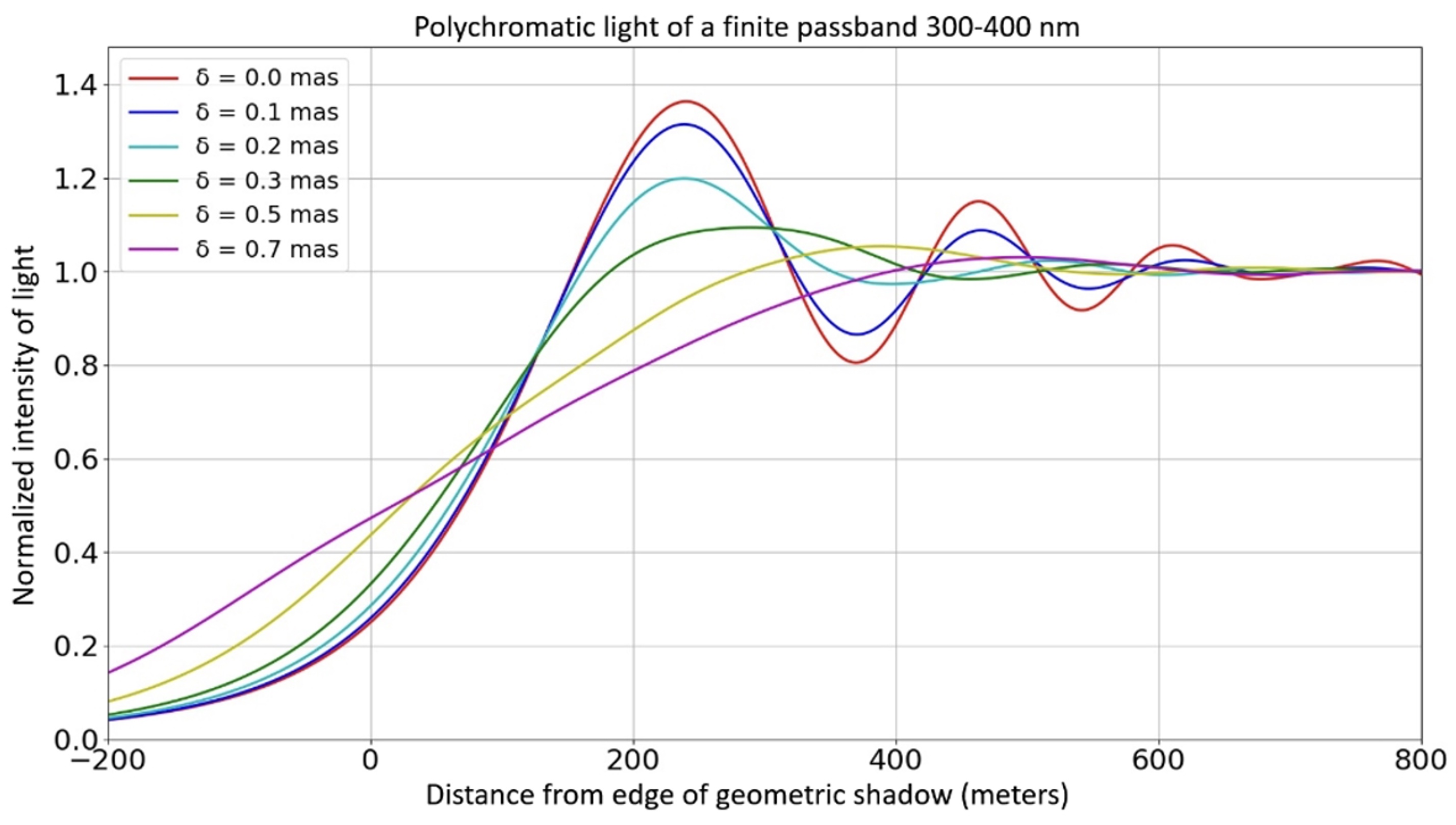}
	\caption{Modeled Fresnel diffraction patterns for a 1.5~AU distant main belt asteroid occultation of a star with varying angular diameters (0.0\,--\,0.7~mas). The intensity profiles show the progressive smoothing of fringes with increasing angular size. Simulations were conducted for a spectral range of 300\,--\,400~nm, highlighting the feasibility of detecting such patterns using high-speed CMOS detectors.}
	\label{fig:asteroid-occult}
\end{figure}

Traditionally, asteroid occultations have been employed to measure asteroid sizes and shapes by timing the duration of the occultation from various observing sites. However, when high-cadence photometry is used, the recorded intensity variations near the occultation edges reveal a rich diffraction pattern (Figure~\ref{fig:asteroid-occult}). For point-like sources, this pattern matches the theoretical Fresnel profile. For extended sources, such as stars, the finite angular size causes a noticeable smoothing and suppression of these fringes. This provides a method to infer stellar angular diameters far below the imaging resolution of even the largest telescopes. This technique has already proven its effectiveness through lunar occultation studies, reaching angular resolutions of about 1~mas, particularly in redder optical wavelengths where lunar background noise is minimized. However, diffraction fitting during asteroid occultations -- thanks to the greater asteroid-to-Earth distances -- can reach sub-milliarcsecond (mas) angular resolutions, approaching 0.1~mas. This surpasses the capabilities of amplitude interferometry \citep{Mozurkewich2003}, which suffers from atmospheric scintillation at blue wavelengths, and intensity interferometry \citep{Hanbury1974}, which is limited to bright, hot stars due to its requirement for very large mirrors. To resolve the fine-scale Fresnel fringes produced during stellar occultations, high temporal resolution—sampling rates of at least 250\,--\,300~Hz is essential. Using the ORCA-Quest\,2 CMOS detector from Hamamatsu, these diffraction patterns can be recorded with enhanced SNRs. Reaching an SNR of 30 or higher is critical for accurately capturing the fringe structure. By fitting the observed light curves to a Fresnel diffraction model, it becomes possible to derive direct measurements of stellar angular diameters. This approach opens a new observational window for investigating the sizes of stars that are too faint or too compact to be resolved through conventional interferometric techniques.

Precise measurements of stellar diameters are fundamental to stellar astrophysics, allowing for stringent tests of stellar structure and evolution models. This is especially critical for low-mass stars, such as M-dwarfs hosting transiting exoplanets, where uncertainties in stellar radii directly propagate into errors on planet size and density \citep{vonBraun2014}. Recent studies have shown that catalogs such as  the K2 Ecliptic Plane Input Catalog (K2 EPIC) tend to misclassify subgiants as dwarfs \citep{Stephens2017}, systematically underestimating stellar radii. This has a significant effect on the inferred properties of planets, particularly those in the Earth-size regime. Thus, direct, model-independent stellar diameter measurements offer a pathway to recalibrating such catalogs and refining exoplanet parameter estimates. This proposal would not only validate a high-precision technique for measuring stellar radii, but could also achieve the smallest angular diameter measurement ever obtained to date.

\subsection{Detecting rings around trans-Neptunian objects and measuring sizes through observations of stellar occultations at high cadence}
\label{subsec:occult-rings}

\begin{figure}
	\centering
	\includegraphics[width=\columnwidth]{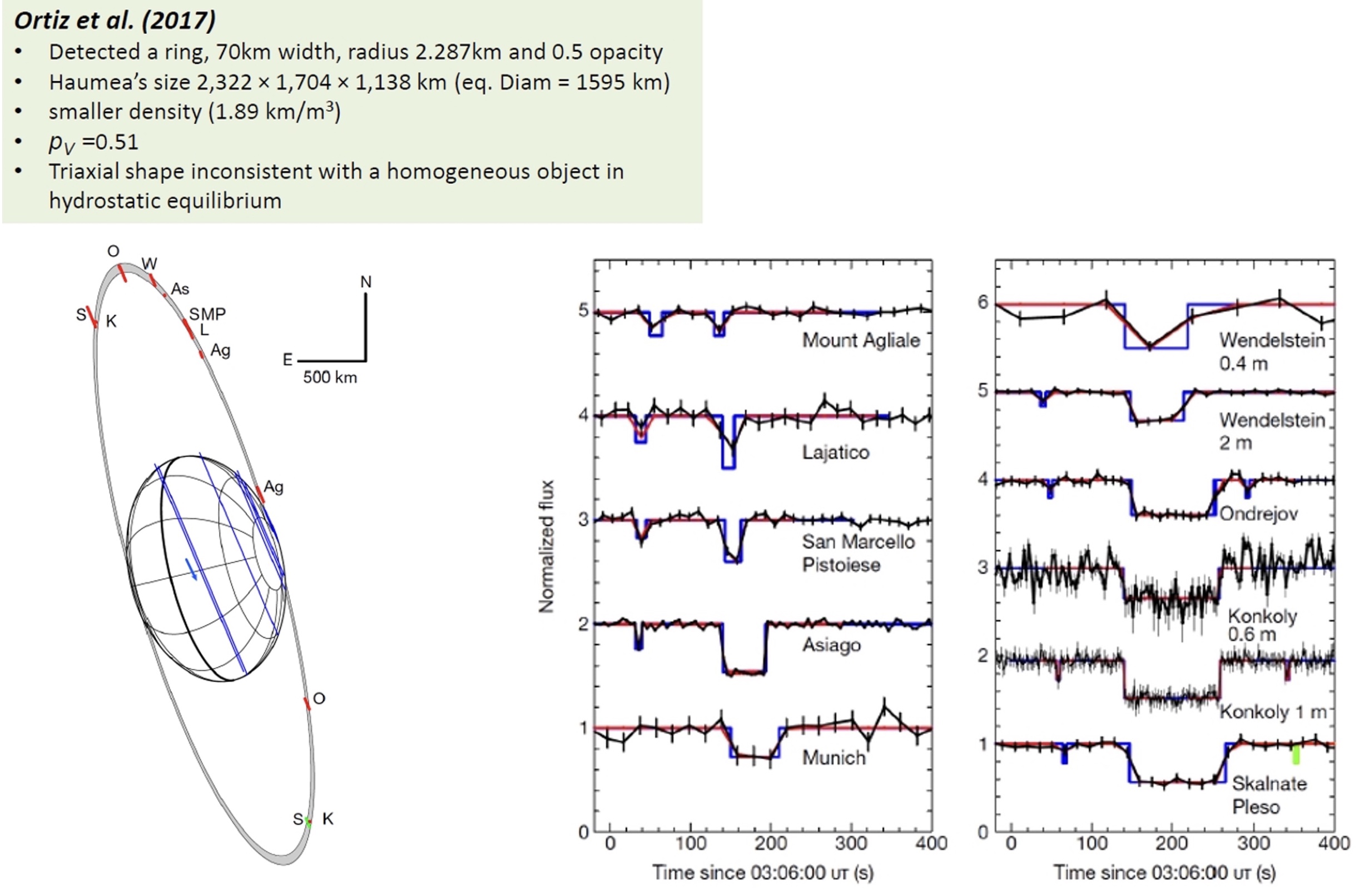}
	\caption{Left panel: Summary of findings from \citet{Ortiz2017} where the discovery of a ring around the TNO Haumea was presented, based on occultation results. The occultation light curves are shown on the right panels, which show stellar flux as a function of time from several observatories. The large drops correspond to the occultation by the main body and the sharp narrow drops correspond to the occultation caused by the ring. The left panel shows the geometry of the ring and the central body, with the occultation chords superposed (in red the ring chords, in blue the body chords).}
    \label{fig:trans-neptune}
\end{figure}

In the last fifteen years, there has been sort of a small revolution in the detailed knowledge of the trans-Neptunian objects (TNOs), thanks to the use of the stellar occultation technique, which allows us to obtain physical properties such as size, shape and geometric albedo with an accuracy that can be only compared to that obtained when a space mission visits a solar system object. Apart from having obtained excellent results in this respect \citep{Sicardy2011,Ortiz2012}, another breakthrough discovery has been the finding of rings around bodies that are not planets (see Figure~\ref{fig:trans-neptune}) such as Haumea, Chiron and Chariklo \citep{Ortiz2015,Ortiz2017,Braga-Ribas2014}. This has been an important discovery with abundant implications, which opens a new field of study and gives rise to many questions, including how frequent these structures are in the current TNO population, what the mechanisms that generate them are, how they can survive for long times, what their compositions are, what the size distributions of particles and fragments that orbit in these structures are. Even a ring system outside the Roche limit has been recently found by \citet{Morgado2023}, with deep consequences. All these extremely novel questions are part of our current research and call for obtaining more and better occultation observations. 

So far rings have only been discovered in four outer solar system bodies, but only around 20 stellar occultations by outer solar system objects have been obtained with high enough temporal resolution and signal to noise to detect rings. It is important to highlight that the ring systems are usually structures of only 10\,--\,20~km of width and given that the typical shadow speed during a TNO occultation is at the 20~km~s$^{-1}$ level this means that the stellar occultations caused by rings are brief phenomena of only around a second. And, on the other hand, if the rings are not optically thick, the dimming of the star light is not comparable to that of an opaque body. For that reason, both high time resolution and high SNR are needed to systematically study occultations by these bodies to systematically look for rings. Typical instruments at most telescopes do not allow fast cadence and typical CCDs have very slow readout times. Only EMCCD cameras and CMOS cameras offer this possibility. At a 2m-class telescope, an instrument such as the ORCA-Quest\,2 qCMOS can allow us to detect stellar occultations of stars as faint as 20~mag with enough SNR to detect rings. 

Thanks to the existence of the GaiaDR3 catalog it is now possible to predict stellar occultations of stars as faint as 20~mag  so we can take advantage of these new recent developments to carry out a specific program on stellar occultations by TNOs.

The discoveries of rings were made with data taken in 2011, 2013, 2017, and 2020 and at the current discovery pace, we may find one or two more ring systems in the next five years. The goal is to boost this discovery rate and find considerably more rings.

\subsection{M-star flares and their effect on exoplanet searches}
\label{subsec:flares}

\begin{figure}[h!]
	\centering
	\includegraphics[width=\columnwidth]{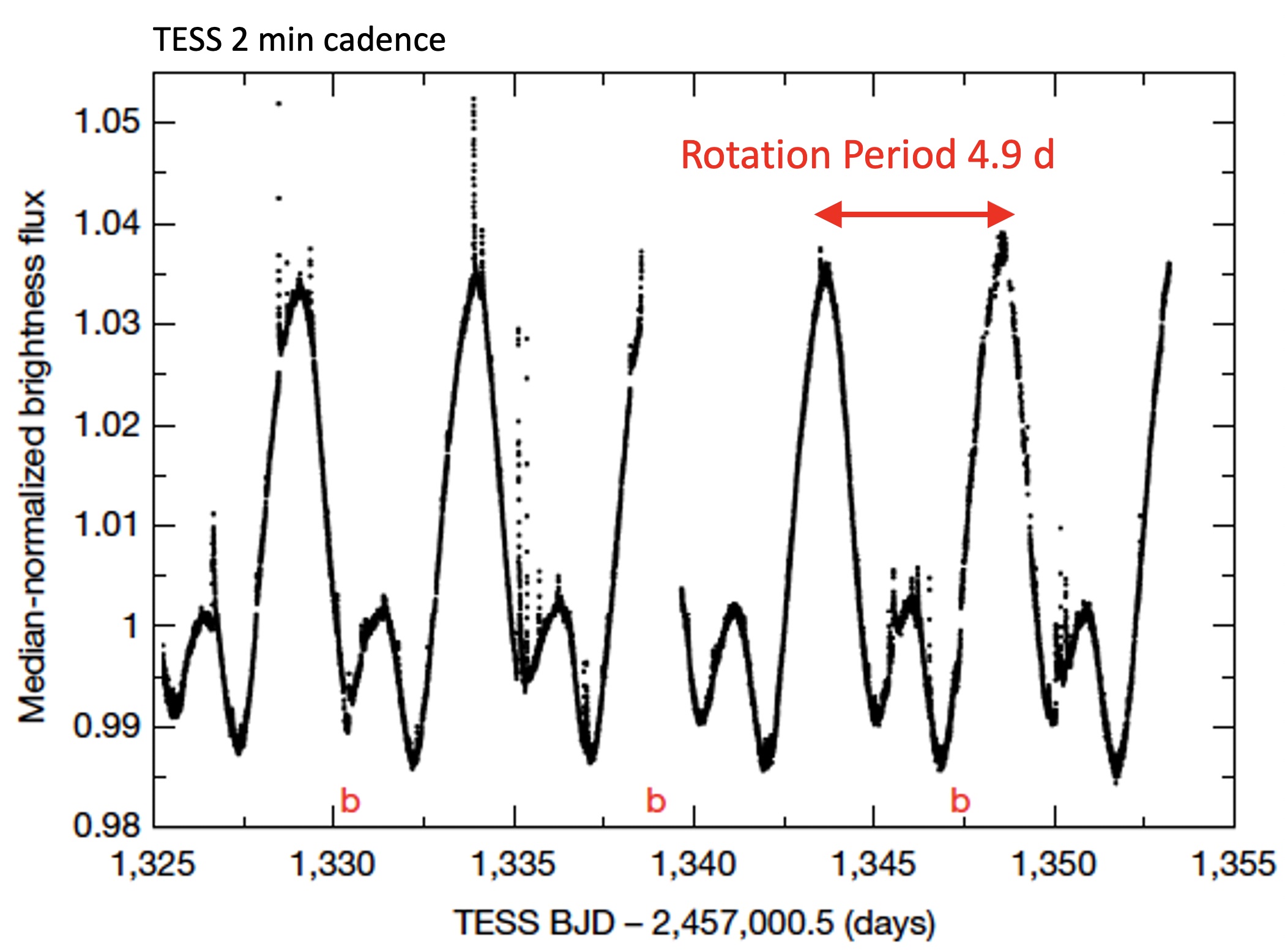}
    \includegraphics[width=\columnwidth]{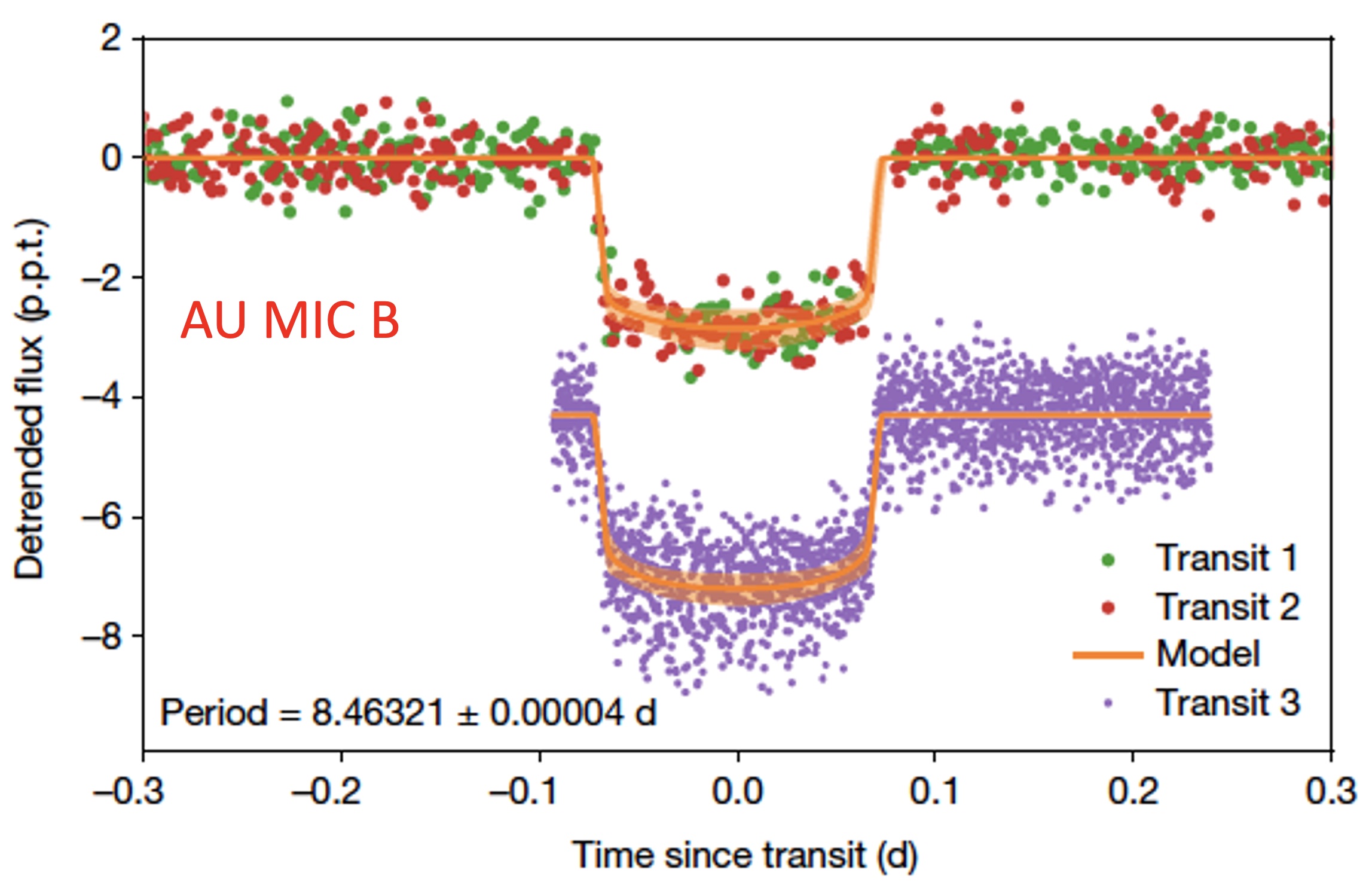}    
	\caption{TESS Light curve of pre-main sequence star AU Microscopii, observed with 2-minute cadence (top). Careful inspection of the normalized light curve (bottom) reveals a transiting exoplanet at phases (b), however severely affected by strong flares. Adapted from \citet{Plavchan2020}.}
	\label{fig:AU_Mic}
\end{figure}

Stellar flares shape the environment around stars and are expected to impact atmospheres of nearby exoplanets dramatically. While solar flares can be studied over many orders of magnitude in terms of the released flare energy and duration, stellar flares are typically only observed at comparatively large flare energies and long durations, due to both brightness and time cadence limitations of available instruments. 

However, since flare occurrence rates follow a power law, smaller flares should happen abundantly on stars of solar mass and lower. Such small flares are accessible with high time-cadence photometry from the ground. Suspected breaks in the flare power law are of high interest to the stellar and exoplanetary science fields, and can be addressed with such data. An additional angle is the selection of suitable filters, particularly in the blue, since flares have better contrast to the quiet photosphere there and that range is underexplored with the current white-red broad-band missions such as Kepler, TESS, and PLATO. 

With the proposed fast qCMOS ORCA-TWIN cameras, we could investigate the occurrence rates and thermal properties of small flares, ranging from sun-like stars to (nearby and therefore bright) M dwarfs, and characterize their impact on exoplanets in close orbits. For the M dwarf targets the investigated small flares may have tangible effects even on the habitable zones around those stars \citep{Tovar2022}.

Moreover, as shown in Fig.~\ref{fig:AU_Mic}, exoplanet transits are badly affected by flares by diluting the exact timing of ingress and egress phases, thus compromising the accuracy of the exoplanet parameter such as orbit and mass. However, masking out flares with high time resolution holds the promise of a major improvement.

\subsection{Pulsations and binarity of hot subdwarfs} 
\label{subsec:sdOB}

Hot subdwarfs (sdO/B stars) are compact, typically helium-burning stars that have lost nearly all of their hydrogen envelopes. Their formation is thought to be dominated by binary evolution, yet the underlying channels and mass-loss mechanisms remain poorly constrained \citep{Heber2026}. Radial-velocity (RV) surveys have shown that a large fraction of sdB stars reside in close binaries, most likely formed through a common-envelope phase \citep{Geier2022}, while light-curve surveys have revealed many additional systems through light variations caused by eclipses, reflection effects, ellipsoidal deformations, and Doppler beaming. The binaries typically have orbital periods from less than one hour to several days \citep{Schaffenroth2019,Schaffenroth2022,Schaffenroth2023_2}. Some of these systems are candidates for Type Ia supernova progenitors \citep{Pelisoli2021}. In addition, many hot subdwarfs exhibit rapid multi-periodic pulsations on timescales of minutes to hours, providing unique constraints on their internal structure \citep{LynasGray2021,Uzundag2024}.

\begin{figure}[h!]
	\centering
	\includegraphics[width=\columnwidth]{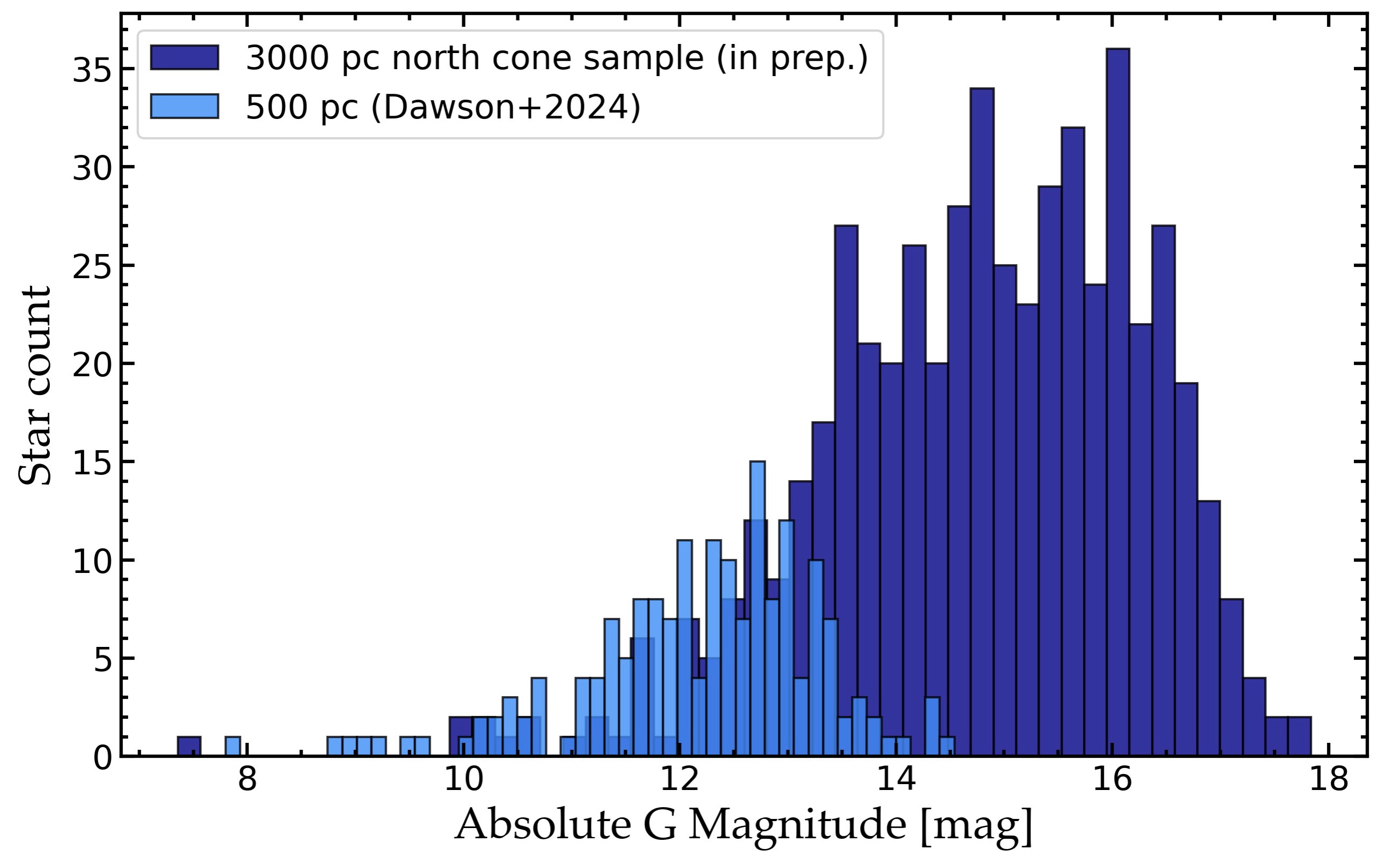} 
	\caption{Distribution of sdO/B-type stars vs.\ absolute G magnitude for two different volume-limited samples.}
	\label{fig:sdOB_stars}
\end{figure}

Earlier estimates of hot subdwarf binary fractions and population properties were strongly affected by selection biases, favouring short periods and large light-curve and RV amplitudes. \textit{Gaia} has transformed this field by enabling to compile homogeneous, astrometry-based samples, ranging from large candidate catalogues \citep{Culpan2022}
to fully spectroscopically classified, volume-limited samples \citep{Dawson2024,Dawson2026}, see Figure~\ref{fig:sdOB_stars}. At the same time, wide-field spectroscopic surveys such as, e.g., the 4-metre Multi-Object Spectroscopic Telescope, 4MOST \citep{deJong2019}, are rapidly increasing the number of confirmed systems requiring detailed time-domain follow-up.

Although photometric surveys (e.g. Gaia, TESS, ZTF) provide invaluable discovery data, they are limited by cadence, exposure time, crowding, pixel size, and bandpass flexibility, particularly for resolving shallow eclipses or detecting low-amplitude, rapid pulsations. The ORCA-Quest2 camera, with its high efficiency and fast readout, is therefore ideally suited for targeted photometric follow-up of spectroscopically identified hot subdwarfs. Mounted on the CAHA 1.23m and Teide Observatory STELLA telescopes, it enables high-cadence, multi-filter observations to resolve eclipses, obtain full orbital phase coverage, and detect weak pulsations on minute timescales. In combination with Gaia-selected, volume-limited samples  (see Fig.~\ref{fig:sdOB_stars} for two samples with different depths) and modern spectroscopic surveys, ORCA-Quest2 provides a powerful and complementary tool to characterize hot subdwarf binaries and pulsators in an unbiased manner.

\subsection{Orbital decay of LISA gravitational wave sources}
\label{subsec:orbitaldecay}

More than 97\% of stars will end their life as white dwarfs (WDs), making WDs by far the largest group of stellar remnants, with many in binary systems. Binaries consisting of two stellar remnants can reach orbital periods as short as a few minutes and physical separations between components smaller than the Earth--Moon distance (e.g.\ \citealt{roelofs+10}). The formation of short period binaries requires two CE phases -- a phase where two stars temporarily orbit within a shared envelope \citep{ivanova+13}. The CE phase is still poorly understood despite it being a critical ingredient for compact binaries, including ultracompact double neutron star and double black holes, which can be detected by LIGO during their merger. For extreme mass ratios the final merger of the two components is prevented, leading to a stable mass transferring ultracompact AM CVn system \citep{marsh+04}. LISA is a space-based gravitational wave (GW) detector acting at lower frequencies than LIGO \citep{colpi+24}. Compact binaries are strong GW sources and will dominate the population of GW emitters in the LISA band allowing for precise multi-messenger studies if we find these sources over the next years \citep{amaro+23}. Systems with orbital periods P$_{orb} < 20$~min will be the strongest and dominant Galactic LISA sources and will be detected by LISA within weeks to months \citep{kupfer+24}. In many cases electromagnetic studies are even required to break degeneracies in LISAs GW data of compact binaries (e.g.\ between masses and distance). In particular, the orbital decay due to GWs encompasses a direct measurement of the chirp mass which is a combination of both masses. However, for almost all sources, LISA’s observing period will not be sufficient to detect the orbital decay which will make the extraction of the chirp mass difficult with LISA data alone. 

Observed minus computed (O--C) studies of compact binaries provide an ideal tool to measure the orbital period due to a shift of the arrival time of eclipses or other photometric (e.g.\ ellipsoidal modulation) variability caused by the binary motion \citep{kepler+91}. For a typical LISA GW binary, the shift in arrival times due to the orbital decay is of the order of at least several seconds over a few years (e.g.\ \citealt{hermes+12,burdge+19,burdge+20}). Precise high-time resolution and high signal to noise ratio photometry with modern CMOS detectors with no readout penalties between exposures are the ideal tool for these studies. 

In a recent study, \citet{Teckenburg2025} obtained high-cadence observations with the 1.23m telescope equipped with the ORCA-Quest~2 qCMOS detector. Combined with data taken with HiPERCAM and the Nordic Optical Telescope (NOT) as well as the 1.2m Oskar Lühning telescope at Hamburg Observatory, \citet{Teckenburg2025} detect an orbital decay of the ultra-compact binary ZTF\,J2130 \citep{Kupfer2020a}. Using the $O-C$ method, they find an orbital decay of $\dot{P}=(-1.48\pm0.27)\times10^{-12}$~s~s$^{-1}$ which is fully consistent with predictions from spectroscopy and light-curve modeling assuming angular momentum loss only from GWs. Most known LISA sources are relatively bright\footnote{See \url{https://gitlab.in2p3.fr/LISA/lisa-verification-binaries} for an overview of the currently known LISA detectable binaries.} and as such this pilot study showed that 1m-class telescopes equipped with modern CMOS detectors with low readout noise and large quantum efficiency are the ideal tool to monitor LISA GW sources to detect the orbital decay over the next few years before LISA’s launch.

\subsection{High cadence photometry of accreting compact white-dwarf binaries} 
\label{subsec:WD-binaries}

Accreting compact WD binaries, aka cataclysmic variables (CVs), are the most common end products of close binary evolution. They consist of a WD primary that accretes matter via Roche-lobe overflow (RLOF) from a low-mass companion (a main sequence star or another degenerate object). More than 90\% of the systems have binary orbital periods between 5\,min and 4\,h and, in very general terms, they evolve under the influence of angular momentum loss (AML) through magnetic braking and gravitational radiation, from long to short orbital periods \citep[see e.g.][]{belloni+schreiber23}. Understanding close binary evolution is of imminent relevance to uncover pathways to supernovae type Ia and to understand the foreground signal of the planned space-based gravitational wave observatory. Yet there are fundamental unsolved problems concerning their sample properties and their evolution \citep[see][and references therein]{ortuzar-garzon+24,schreiber+24, pala+25}.

High-speed photometric observations are one but essential step to address unsolved questions about (a) the intrinsic orbital period distribution and thus the strength of angular momentum loss, about (b) the mass distribution of the WDs along their evolutionary sequence, about (c) the spin periods and the spin period evolution of the accreting WDs (between minutes and hours), and (d) about period changes which could be mediated by unknown angular momentum loss mechanisms or the occurrence of circumbinary planets. 

While the demands on time resolution are moderate (of order minutes) to just measure orbital or spin periods (cases (a) and (c) above), one needs data with high time resolution (about 1\,s) to measure WD radii and orbital period changes (cases (b) and (d) above). However, given the typical rather low brightness of potential targets, even for observations with moderate requirements on time resolution, an sCMOS camera is preferred over a conventional CCD due to its excellent noise properties and the negligible dead time at the same QE. 

\begin{figure}[h!]
	\centering
	\includegraphics[width=\columnwidth]{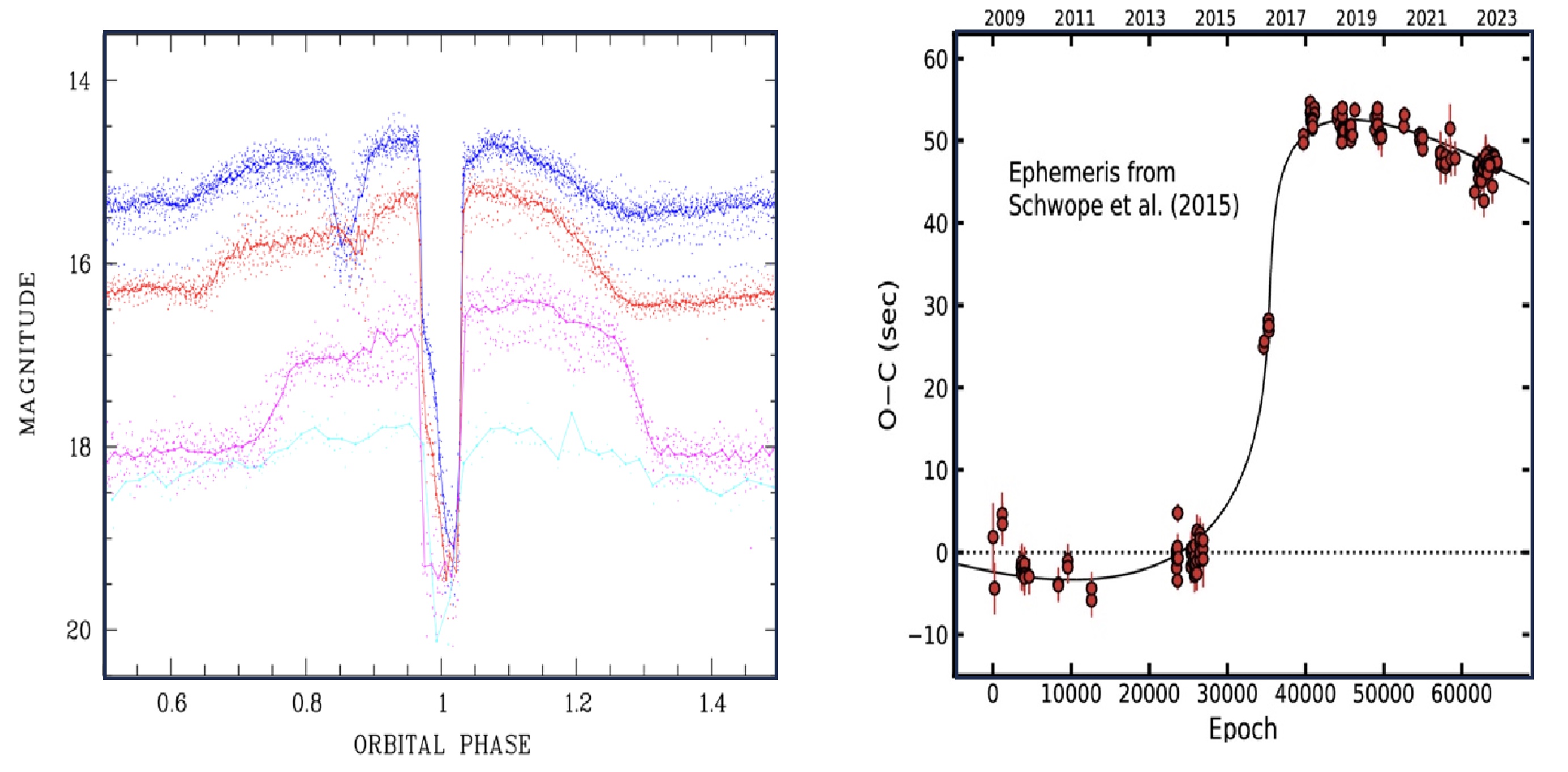} 
	\caption{Left panel: Photometry of the eclipsing polar V808 Aur \citep{Schwope2015}. The eclipse with a depth of up to 5~mag allowed to establish a precise long-term ephemeris. Right panel: sCMOS observations with a 1-second time resolution allowed to resolve the eclipse. Eclipse timing variations with a strong gradient in 2016 were interpreted as due to a circumbinary hot Jupiter in an 11-year elliptical orbit \citep{Leichty2024}.}
	\label{fig:WD_binaries}
\end{figure}

A direct measurement of the radius of a WD is possible through detailed eclipse observations and adapted models. The mass of the WD is then inferred within small margins via a mass--radius relation \citep{bedard+20}. Eclipse ingress and egress typically last less than about 30~s, and these events need to be well resolved. Substructures on the surface of the WD-like accretion-heated or cyclotron emitting hot spots will be uncovered and need to be included into the modeling process for ultimate precision \citep{schwope+26}, see Figure~\ref{fig:WD_binaries}. 

Long-term monitoring of magnetic CVs have uncovered considerable period changes through precise eclipse timing. In two cases, the period changes were interpreted as light travel time (LTT) effect caused by circumbinary planets \citep{beuermann+11, Leichty2024}. Other CVs have shown a much more complex timing pattern which could not yet be convincingly explained \citep{bours+14, bours+16}. 

Currently, we are facing a steeply growing CV population through the spectroscopic identification of Gaia and eROSITA candidates \citep{salvato+25, schwope+24b}. Soon complete volume-limited samples of the various subclasses of CVs comprising several hundred objects will be available. A bottleneck to fully characterize such samples is the determination of the orbital periods of their members. Dedicated facilities with ORCA-Quest\,2 cameras are instrumental to study the timing behavior of the members, to derive their underlying spin and orbital periods and uncover eclipse timing variations.

\subsection{Search for hypothetical primordial black holes in the solar system}
\label{subsec:PBH}

Primordial black holes (PBH) have been invoked as possible constituents of dark matter (DM), potentially accounting for all of the DM content of the Universe, thus eliminating the need to search for DM particles that have as yet not been found in particle physics experiments. \citet{Carr2025} have noted that the scientific interest in PBH in terms of number of publications has experienced exponential growth over the past 20 years. An extensive review about the formation, properties, and observational constraints of PBH is given by \citet{Escriva2024}.

There are arguments that the possible mass of such PBHs must be in the range of $10^{17}$\,--\,$10^{23}$~g, which is the typical mass of an asteroid. Given such deliberations, it is obvious to consider observational tests that can prove or disprove the existence of such objects. \citet{Cuadrat2024} have proposed to measure the kinematic impact from the passage of PBHs through the solar system using satellite navigation systems. \citet{Tran2024} discuss the use of radar distance measurements between Earth and another planet over a decade to reveal the impulse velocity impact of a PBH passage as a subtle change of orbital phase of the planet concerned. The effect would be very small. For a review of the current discussion see \citet{Thoss2025} who have disputed that such effects on planets would be measurable at all.

\begin{figure}[h!]
	\centering
	\includegraphics[width=\columnwidth]{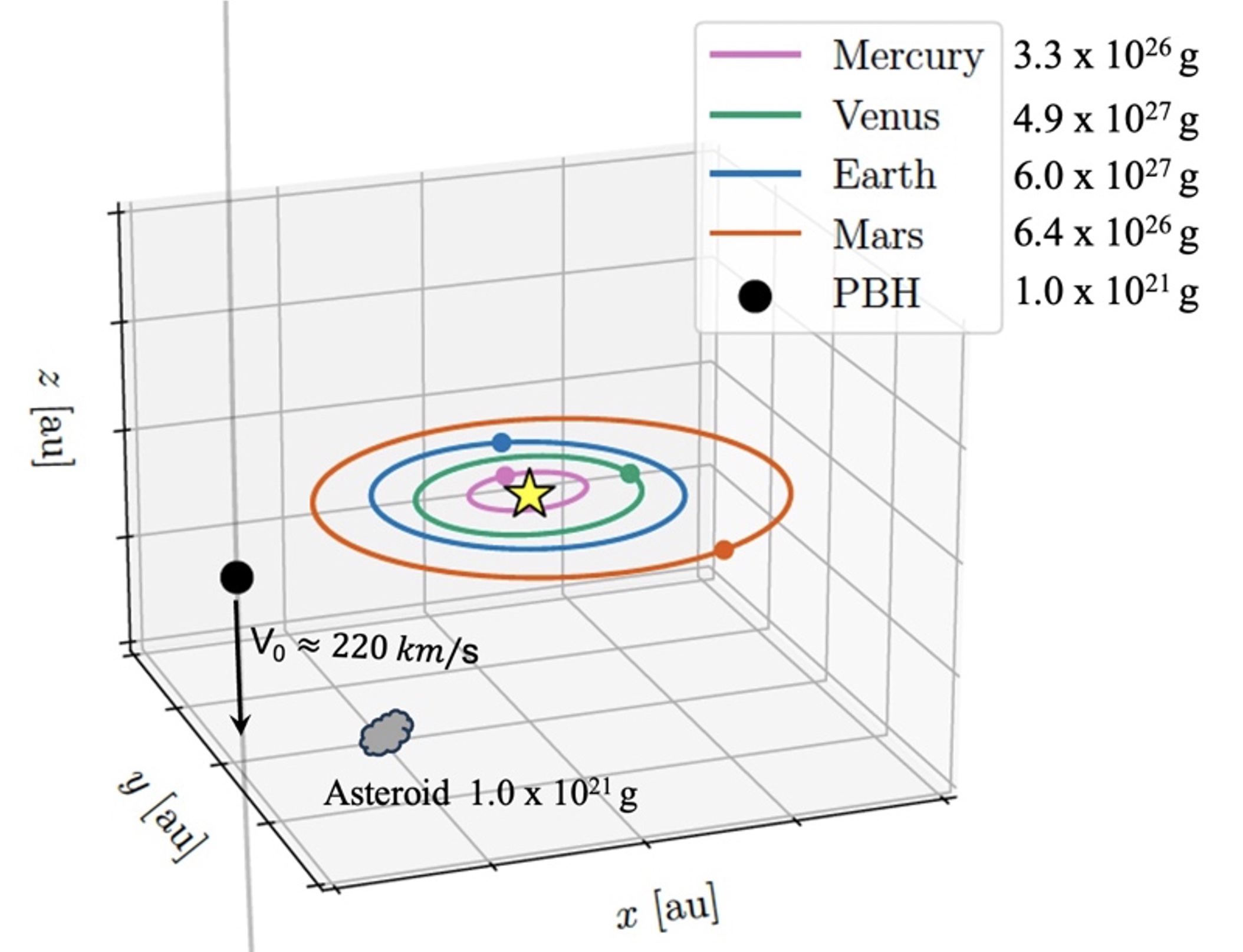} 
	\caption{Solar system with inner planets Mercury, Venus, Earth, and Mars, and the passage of a PBH \citep[adapted from][]{Tran2024}. A close encounter with an asteroid belt object can potentially induce a much stronger perturbation than the ones hypothesized for a six orders of magnitude more massive planet.}
	\label{fig:PBH_asteroid}
\end{figure}

Given the inertia of a planet in comparison to the impact of an asteroid-mass PBH on the one hand, but the comparable mass between an asteroid proper and a PBH, also considering the existing large amount of data for small solar system bodies on the other hand, one can ask the question whether a PBH fly-by would be detectable as a perturbation of the orbit of one or several asteroids that could be tracked with high precision from the ground or in space (Figure~\ref{fig:PBH_asteroid}). The goal of an ORCA-TWIN experiment on PBH would be to assess the accuracy to which asteroid orbits can be constrained in real-time as well as with multi-epoch observations from the ground, and whether this accuracy is good enough to detect the passage of an asteroid-mass PBH through the Solar system. With the anticipated number of several million main-belt asteroids that will be detected by the Vera Rubin Observatory \citep{Rozek2022}, thus the possibility to create complex dynamical models of the Solar System, the opportunity may arise to launch PBH passage alerts from measured asteroid orbit perturbations from such a model. At present, the feasibility of such detections is speculative as the accuracy of known orbits from the model is unknown, and whether a monitoring capability for small perturbations will be available at all. However, the lightcurve analysis from the first 2103 asteroid detections during only nine nights of LSST camera commissioning presented by \citet{Greenstreet2026} is already encouraging enough. Given the 5$\sigma$ depth of $\sim 23\ldots25$~mag for these detections, any follow-up observations would require a larger telescope than available for ORCA-TWIN. Nevertheless, quantitative results from the science cases presented in Subsections~\ref{subsec:triangulation} and \ref{subsec:fastrotNEO} will inform about any advance to be expected from high cadence qCMOS observations to this end.

\subsection{Fast-cadence high-contrast Bayesian imaging with photon counting camera}
\label{subsec:bayesian-imaging}

\begin{figure}[h!]
	\centering
	\includegraphics[width=\columnwidth]{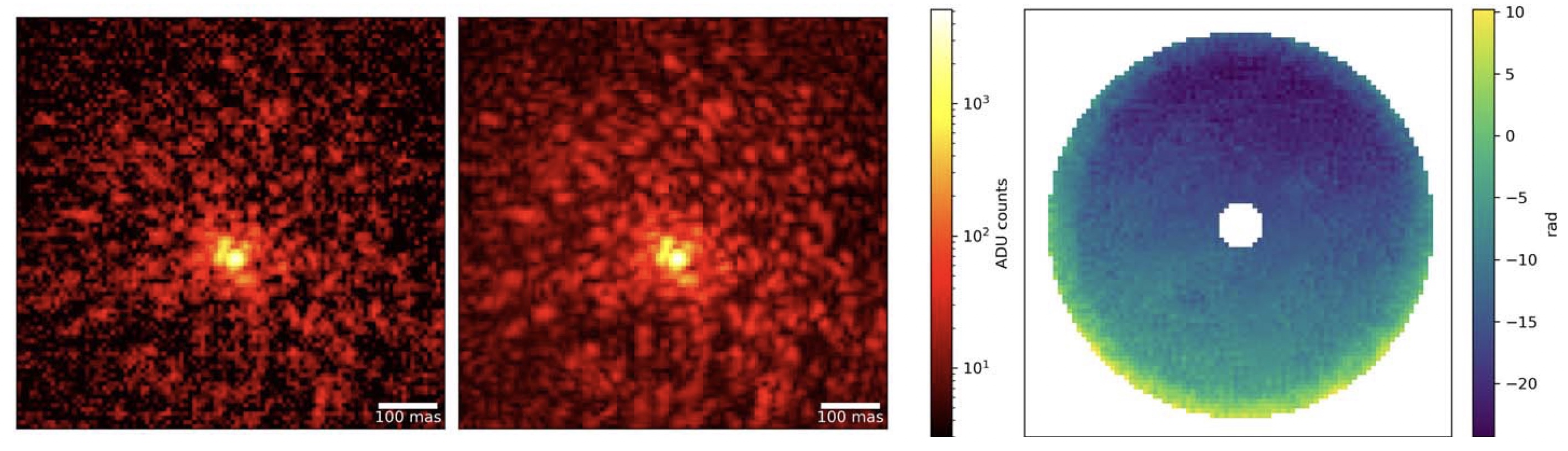}  
	\caption{Left panel: snapshot data (instantaneous PSF convolved star plus shot noise) in the focal plane (rebinned to $100 \times 100$ pixels) of the 5.5~mag star Gl\,777, acquired with the SHARK-VIS Forerunner instrument at the LBT in a 1-millisecond exposure; the interference pattern shaped by the atmospheric fluctuations (``speckles'') is well visible all around the star. Middle panel: HCBI reconstruction of the PSF convolved star (without shot noise) for the corresponding frame, shown with the same logarithmic color map. Right panel: the reconstruction in the pupil plane of the optical path difference field for that frame of the Gl\,777 sequence, which determines the reconstructed PSF shown in the middle panel \citep[from][]{Rothetal2023}.}
	\label{fig:Gl777}
\end{figure}

Direct imaging of exoplanets is challenging due to the high brightness contrast between a star and its exoplanets at very small angular separations, just a few times wider than the largest telescopes' diffraction limit, aggravated by atmospheric turbulence that distorts the incoming wavefront. While extreme adaptive optics (ExAO) partially compensates for these distortions, residual spatiotemporal variations create a complex speckle pattern, changing on a few millisecond timescale, that obscures fainter companions (left panel in Fig.~\ref{fig:Gl777}). To enhance contrast, data post-processing techniques, such as Angular Differential Imaging (ADI) \citep{Marois2006}, Speckle-Free ADI (SFADI) \citep{LiCausi2017}, Local Combination of Images (LOCI) \citep{Lafreniere2007}, Principal Component Analysis ADI (PCA-ADI) \citep{Amara2012,Soumer2012}, leverage field rotation to model and subtract the central star's light and speckles, mitigating residual noise and improving detection capabilities.

Recently, we introduced a new method, dubbed Hight Contrast Bayesian Imaging \citep[HCBI,][]{Rothetal2023} to improve exoplanet detection in high-contrast imaging by exploiting the statistical properties of the rapid variation of the point-spread function (PSF) acquired by the fast-cadence SHARK-VIS camera \citep{Pedichini2024} at the Large Binocular Telescope (LBT), thanks to an explicit forward data modeling that simulates the natural information flow from the signal to the detector. The algorithm is based on the mathematical methods of Information Field Theory \citep[IFT,][]{Ensslin2019} and uses Bayesian inference to simultaneously reconstruct the true object and the PSF evolution from a kilohertz-rate frame sequence acquisition, where the speckle pattern can be considered frozen in each frame. In practice, it infers the instantaneous wavefront at the telescope pupil which produced the observed speckles for each frame (Fig.~\ref{fig:Gl777}), leveraging on a set of prior information reflecting the physics of the object, the atmospheric perturbations, the telescope and detector response, and the photon and readout noise statistics.

In these short acquisitions the individual speckles are imaged in a photon starved regime, a situation where the photon count accuracy would be greatly improved if a photon counting device was used. In this context, a low-noise qCMOS camera such as the Hamamatsu ORCA-Quest2 would provide a noticeable advance with respect to the current Zyla sCMOS sensor on board SHARK-VIS, whose readout noise is larger than the single photon signal which does not allow exact photon counting and reduces the performance of the Bayesian reconstruction. Such a photon counting camera can acquire frames at a rate of 1~kHz in a 256$\times$256-pixel format, which is the standard frame size in SHARK-VIS for close companions targets and, being only limited by photon noise, its usage could reveal the physically faintest detectable exoplanets given the star magnitude, the telescope size, and the atmospheric conditions.

\subsection{Identification of electromagnetic counterparts of gravitational wave mergers}
\label{subsec:GWmergers}

Multi-messenger observations of neutron star binary mergers allow for critical measurements spanning cosmology, nuclear astrophysics, stellar evolution and high-energy astrophysics. To date, however, there has only been one binary neutron star merger, dubbed GW170817 after its discovery date, observed in both gravitational waves, as well as across the electromagnetic (EM) spectrum, revealing two EM counterparts: a kilonova and a short gamma ray burst \citep[e.g.,][]{2017ApJ...848L..12A}. This limited sample of one event reflects not only the lower-than-anticipated binary neutron-star merger rate after GW170817, but also the substantial observational challenges associated with rapidly and robustly identifying EM counterparts within large GW localisation regions. Current GW localisations typically cover tens to thousands of square degrees \citep{2025arXiv250818082T}, while the optical counterparts are faint, fast-evolving over the hours to days timescale, and embedded in a dense fog of unrelated astrophysical transients and variable stars \citep[e.g.,][]{Kasliwal_2020}. End-to-end simulations of GW–EM follow-up demonstrate that successful identification requires a combination of low-latency response, efficient tiling strategies informed by the full three-dimensional GW localisation (sky position and distance), and rapid discrimination of astrophysical false positives \citep{Nissanke2013}. These studies and more recent efforts also show that identifying the UVOIR counterpart is often limited not by sensitivity alone, but by observational efficiency, cadence, and the ability to characterize candidate evolution on short timescales. The first hours to days following a merger are especially rich in information about its physics  \citep[e.g.,][]{Metzger_2019}. Observations of GW170817 revealed a rapidly evolving early UV/optical component whose precise physical origin remains unknown \citep[e.g.,][]{2018ApJ...855..103P,Kasliwal_2017}. Detailed modelling and observational case studies demonstrate that early-time observations strongly constrain competing emission scenarios and that optical data play a crucial role, especially when combined with earlier ultraviolet and later infrared measurements \citep{Dorsman2023}. Capturing this phase requires instruments such as ORCA-TWIN capable of responding within hours, operating efficiently at short exposure times, and revisiting candidates frequently enough to measure rapid colour and luminosity changes that distinguish kilonovae from common contaminants such as supernovae or stellar flares. In concert with other UVOIR facilities in a coordinated manner, ORCA-TWIN enables extended temporal coverage across night–day boundaries, improves resilience against weather, and allows near simultaneous or rapidly sequential observations that strengthen candidate validation. This architecture is particularly advantageous during the crucial first night after a GW trigger, when uninterrupted coverage can determine whether a candidate’s evolution is consistent with a kilonova and/or the optical afterglow of a short gamma-ray burst.

Once an EM counterpart is identified, the scientific return is maximized through coordinated, multi-wavelength observations, particularly in concert with ULTRASAT that may observe the earlier UV emission \citep{Dorsman2023}. Dense early time optical photometry from ORCA-TWIN, combined with contemporaneous or earlier UV observations from ULTRASAT, provides a powerful and perhaps the only discriminant on the origin of the rapidly evolving early ``blue'' emission, enabling distinguishing competing models and tighter constraints on the composition, geometry, and energetics of the ejecta than either wavelength range alone. Secure optical localization further allows for robust association with a host galaxy, enabling environmental and population studies and providing the redshift required for standard siren cosmology when combined with GW distance measures \citep{H02017}. Most critically, rapid and well-vetted optical identification acts as the first step for global ground and space-based follow-up campaigns, triggering spectroscopy to probe composition and velocities, infrared observations to track the emergence of lanthanide-rich ejecta, and longer-term radio and X-ray monitoring to constrain relativistic outflows and jet–cocoon physics. 

\subsection{High-resolution solar observations}
\label{subsec:solar}

The 1.5m~GREGOR solar telescope \citep{Schmidt2012} and the 0.7m~Vacuum Tower Telescope \citep[VTT,][]{vonderLuehe1998} are located within sight of the twin STELLA telescopes at the Teide Observatory. This provides an opportunity to conduct a technical observing campaign to assess the performance of the high-speed (up to 120~Hz) ORCA-Quest\,2 camera for solar high-resolution imaging and spectroscopy before the camera will be installed at STELLA. 

\begin{figure}[h!]
\centering
\includegraphics[width=\columnwidth]{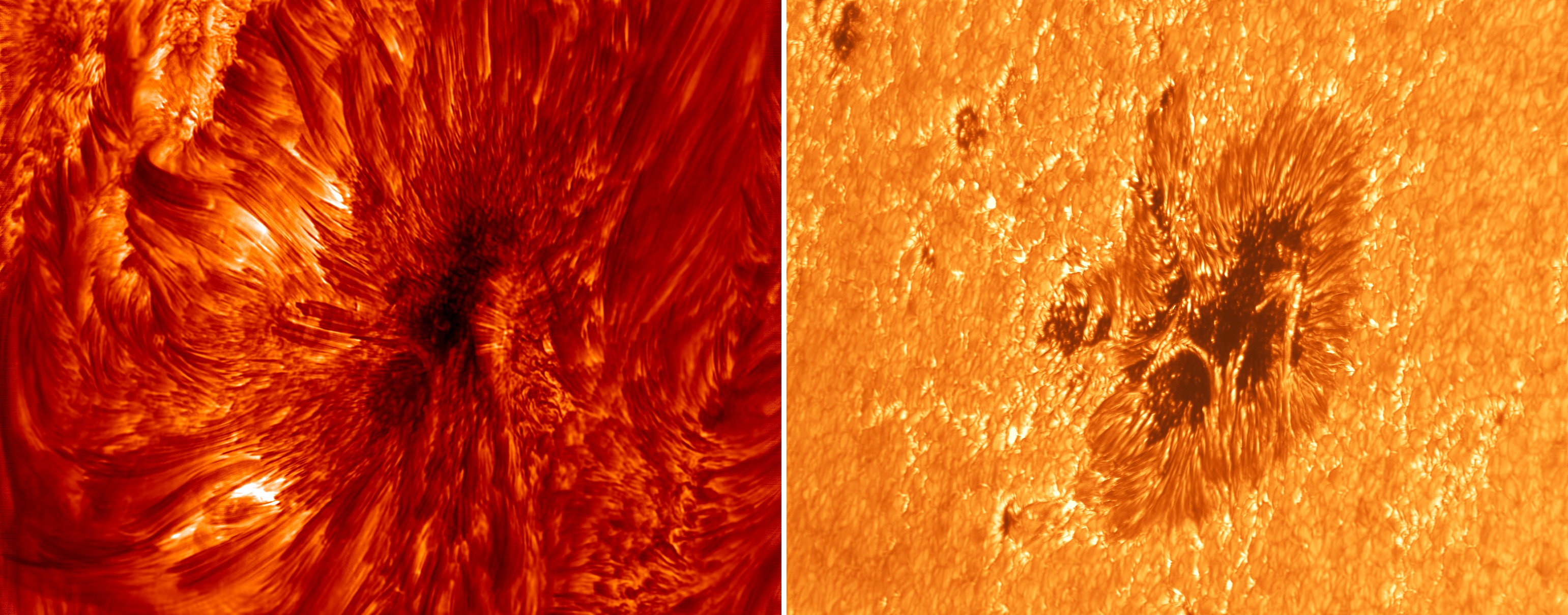}  
\caption{Speckle-restored GREGOR/HiFI+ H$\alpha$ narrow-band (FWHM = 0.6~\AA, left) and 
    broad-band (FWHM = 7.5~\AA, right) images show the leading sunspot in active region NOAA~14149 that was observed on July~20, 2025. The field of view is 76.5\,arcsec $\times$ 60.5\,arcsec.}
\label{HIGHRES}
\end{figure}

The camera will be tested using the improved High-resolution Fast Imager \citep[HiFI+,][]{Denker2023} optical setup at GREGOR. There, the camera can be placed easily in the Ca\,II\,H and H$\alpha$ channels without adapting the plate scale. The photon flux in the line cores of these strong chromospheric absorption lines is very low \citep[e.g.,][]{Sobotka2016}, especially when observing sunspot umbrae and using low-transmission, narrow-band filters. In addition, short exposure times ($t_{exp} < 10$~ms) are necessary for image restoration using speckle \citep{Woeger2008} and blind deconvolution \citep{Loefdahl2002} techniques. An example of speckle-restored HiFI+ H$\alpha$ narrow- and broad-band images is shown in Fig.~\ref{HIGHRES}. Furthermore, the acquired data will enable a comparison of the qCMOS camera with the HiFI+ CMOS and sCMOS cameras, which use Sony Pregius IMX~174 and Fairchild Imaging CIS2051 detectors, respectively. Data reduction is straightforward using the `sTools' IDL data processing pipeline \citep{Kuckein2017, Denker2018}, as only the routines for importing data need adaptation. 

Another means of assessing the performance of the ORCA-Quest\,2 is the VTT echelle spectrograph, to which the camera can be mounted to record drift scans across the solar disk \citep[see][for details]{Verma2025}. These observations will also enable us to compare the ORCA-Quest\,2 camera with the large-format CMOS detectors (CMV50000-1E3M1PA) of the Fast Multi-Line Universal Spectrograph (FaMuLUS) camera system at the VTT echelle spectrograph.

Based on manufacture information and test results, the ORCA-Quest\,2 qCMOS camera has the potential to become the ``work horse'' for imaging, spectroscopy, and polarimetry at 1-meter-class solar telescopes. The largest impacts on science are expected for the study of the umbra--penumbra interface with penumbral grains, umbral dots, and light bridges, where the interaction of small-scale magnetic fields and convection is still not sufficiently understood. The low-noise and high-dynamic-range properties also promise advances in understanding highly dynamic and at times eruptive prominences, filaments, and flares. During the ``solar intermezzo'', within the ORCA-TWIN project, related science cases will be selected for the GREGOR and VTT observing campaign, aiming at science-ready data.

\section{qCMOS image sensor technology} \label{sec:qcmos}

On October 17, 2024, two Hamamatsu ORCA-Quest\,2 model C15550-22UP cameras were delivered to DZA, with setup and initial testing beginning in the newly established DZA detector laboratory in Görlitz. This new model distinguishes itself from the first generation ORCA-Quest as described by \citet{Lucas2024} and \citet{Strakhov2024} in that the QE in the UV was significantly improved through an optimized anti-reflective coating on the camera window, and a factor five improvement of frame rate (25~Hz) for the low readout noise mode. It has been demonstrated that the low readout noise of 0.3~e$^-$ enables detection of single photons, thus literally resolving the discrete Poissonian distribution of photon detections in the very low light level regime.

This achievement has been made possible by the optimization of the CMOS sensor chip architecture featuring an individual output node transistor for each pixel, and a multiplexer that connects the charge-to-voltage converted pixel signals to the outside circuitry. 

\begin{figure}[t!]
	\centering
	\includegraphics[width=\columnwidth]{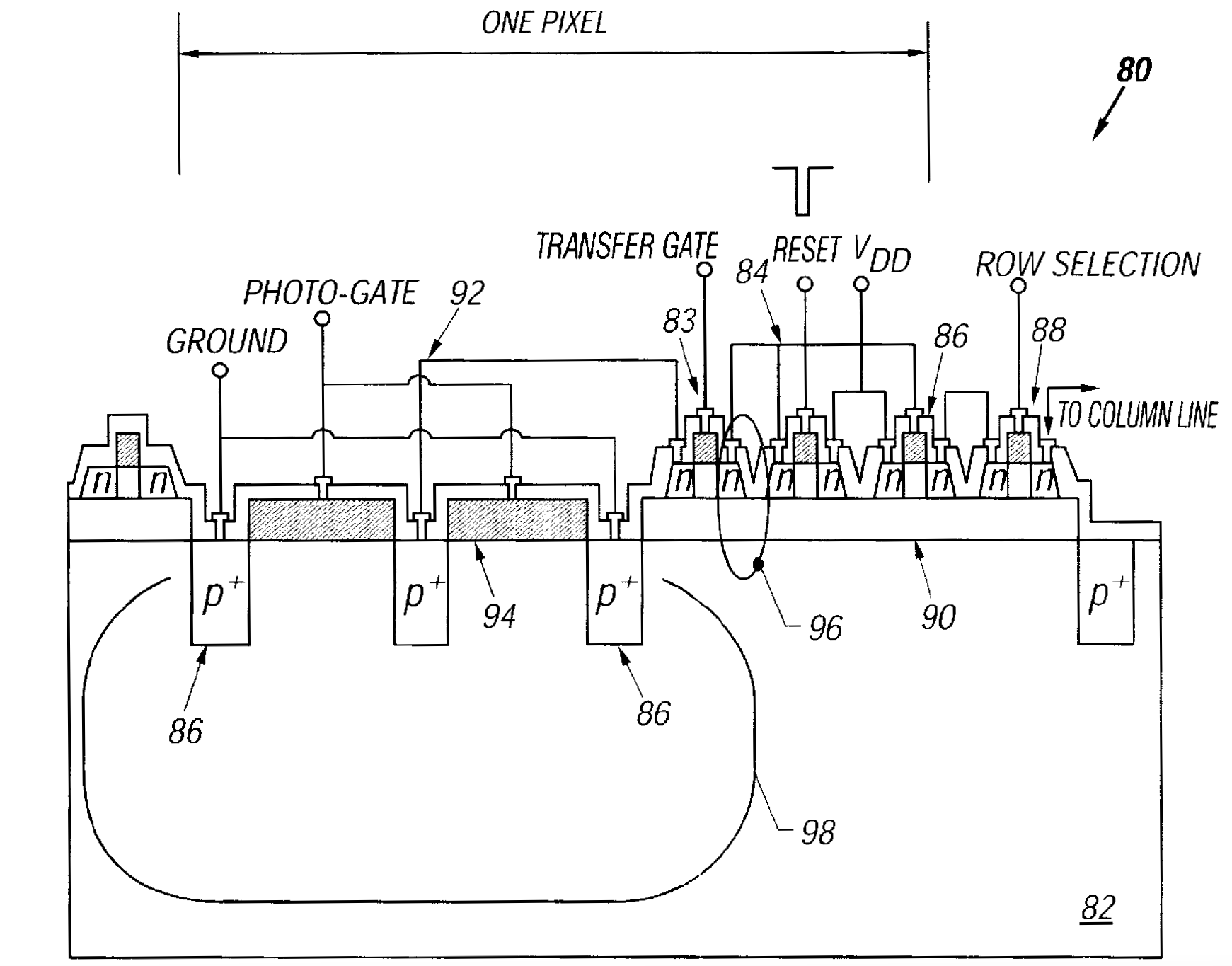}
	\caption{CMOS pixel layout, including the floating transfer gate detail. Legend: pixel (80), thick low-doped p-type silicon substrate (82), transistors (83), (84), (86), and (88), buried oxide insulator layer (90), photodetector area (82), polysilicon gates defining photodetector (94), deep depletion region for photo electron collection (98), reset transistor (84),  buffer transistor (86), row selection transistor (88), transfer gate transistor (83), and floating sense node (96). Reproduced from \citet{Pain2002}, US Patent No.\ US 6,380,572 B1.}
	\label{fig:floating-TrGate}
\end{figure}

As an example, a schematic picture of a CMOS sensor pixel architecture is shown in Fig.~\ref{fig:floating-TrGate}. A relevant technical detail of this layout is the {\em floating diffusion node} (FDN), item~(96) in Fig.~\ref{fig:floating-TrGate}. It is a small, electrically isolated region of silicon that plays an important role for the conversion of photocharge to a measurable voltage. As dictated by the geometry of the node, it has a certain capacitance $C$ that gives rise to a voltage $U$ when it is loaded with a photocharge $Q$ according to the elementary equation $U = Q/C$. Because of the tiny geometry of the FDN and its separation from the photo-collection area (98), the capacitance is very small, hence the voltage per photocharge (conversion gain) is high. This voltage controls the output current of the amplifier, allowing the small, sensitive voltage signal to be buffered and amplified before being read out from the pixel. The FDN is ``floating'' because it is not connected to a fixed or controlled voltage source during the charge integration and sensing period. The reset transistor periodically discharges (resets) it to a reference voltage before a new charge integration period begins which allows to implement the correlated double sampling (CDS) noise reduction technique \citep{Janesick2001}. Unlike the earlier realization of CDS for CCDs which happened in an external video card of the controller electronics, CMOS CDS is implemented on-chip, thus avoiding external wiring and the associated noise sources.

It is the optimization of conversion gain and CDS that enables the Hamamatsu qCMOS sensor to deliver the extremely low readout noise figure of 0.3~e$^-$. As pointed out by the manufacturer, read noise depends on the square root of the bandwidth. Therefore, each column of the CMOS image sensor is supplied with its own low-pass filter circuit, so that the time needed to read out a complete line can be employed for low-pass filtering of the pixel signal selected by the pixel switch, thus reducing the readout noise even further.

A detailed study and characterization of the first generation ORCA-Quest camera was presented by  \citet{Lucas2024}. It includes a performance comparison between the ORCA-Quest upgrade and the EMCCD camera previously used in the VAMPIRES instrument within SCExAO on the Subaru telescope. The authors conclude that the qCMOS detector performs better than the EMCCD over a broad range of illuminations, especially at low-light levels.

\citet{Krynski2025a} have characterized the  ORCA-Quest\,2 camera C15550-22UP for space applications, also comparing to the EMCCD. They find that for the same illumination level the qCMOS sensor has an advantage over EMCCD in terms of SNR due to high conversion gain as opposed to the multiplication mechanism of the EMCCD, where the amplification of noise explains the limiting factor at low fluxes. The low dynamic range of EMCCDs is prompting another advantage of the qCMOS sensor. More recently, the use of the BAE HWK 4123 sensor was proposed for the Lazuli Space Observatory \citep{Roy2026}.

\section{Numerical simulations} \label{sec:simulations}

In order to understand in how far the ORCA-Quest\,2 camera is competitive on sky with regard to high performance CCD cameras, we have conducted a performance comparison between two types of commercially available sensors. The following systems that are available on the market were taken as reference and used with parameters listed in their respective data sheets: the Hamamatsu ORCA-Quest\,2 camera using the $4096\times2304$ / 4.6~$\mu$m pixel qCMOS sensor, and the Andor iKon-L using the $2048\times2048$ / 13.5~$\mu$m pixel Teledyne e2v CCD42-40. The main point is to quantitatively investigate the advantage from low read noise, despite the small pixel size. For the reasons explained in Sect.~\ref{sec:qcmos}, we have passed on to include EMCCDs in the simulations as well.

The parameters used for the two camera types were taken from the manufacturer's ORCA-Quest\,2 C15550-22UP datasheet,\footnote{\url{https://www.hamamatsu.com/content/dam/hamamatsu-photonics/sites/documents/99_SALES_LIBRARY/sys/SCAS0166E_C15550-22UP.pdf}} ORCA-Quest\,2 White Paper,\footnote{\url{https://www.hamamatsu.com/content/dam/hamamatsu-photonics/sites/documents/99_SALES_LIBRARY/sys/SCAS0149E_qCMOS_whitepaper.pdf}} iKon-L 936 CCD datasheet\footnote{\url{https://andor.oxinst.com/products/ikon-large-ccd-series/ikon-l-936}} and are listed in the Appendix.

\begin{figure}[h]
	\centering
	\includegraphics[width=\columnwidth]{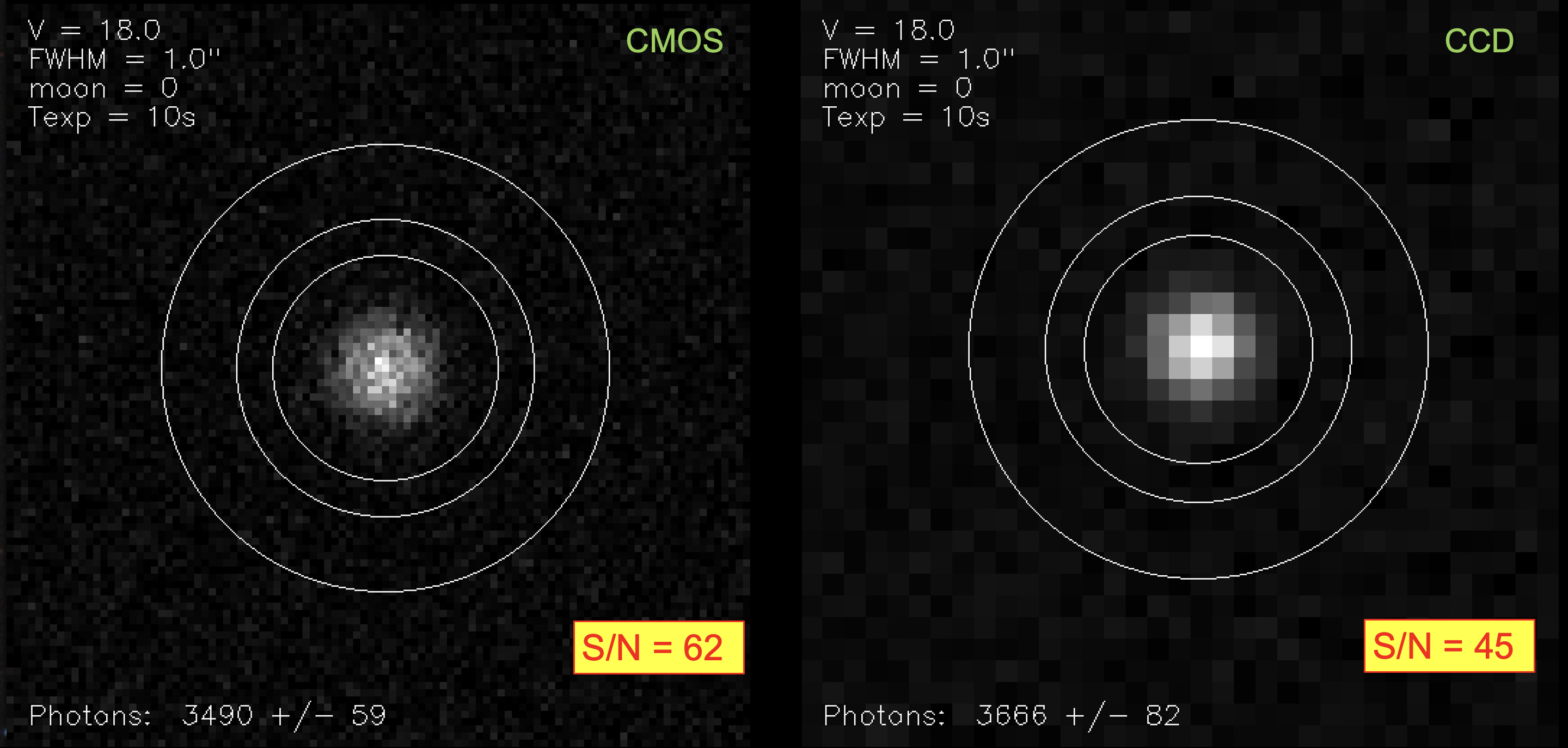}
	\caption{Simulated images of a star for qCMOS (left) and CCD sensor (right), respectively, within a window of approximately $10\,arcsec \times 10\,arcsec$ width. The concentric circles indicate the aperture and the sky annulus for DAOPHOT photometry that is applied to the images to measure photon flux and its uncertainty.}
	\label{fig:V18_10s}
\end{figure}

The simulation generates images of a star in a window of approximately 10\,arcsec $\times$ 10\,arcsec width at the pixel sampling given by the qCMOS and CCD sensors, respectively, at the plate scale of 20.9\,arcsec\ mm$^{-1}$ for a 1.23m telescope with a focal length of 9.8~m and a diameter of 0.58~m of the central obstruction. The photon flux for a star of given magnitude in a particular filter is computed from tabulated values for filters U, B, V, R, I, u, g, r, and z, and multiplied with correction factors for the light collecting area of the telescope, for atmospheric extinction, instrumental throughput, and quantum efficiency at the respective wavelengths, see Table~A.1 in the Appendix. 
 
The PSF is modeled with a Gaussian, scaled to the total stellar flux in units of photon counts for a given exposure time, the FWHM of atmospheric seeing, and then sampled at the plate scale mentioned above. Likewise, the sky background is estimated from tabulated sky surface brightness values at the wavelength of a given filter for Moon phases between new and full moon (see Appendix). The uncertainty of resulting photon counts in each pixel is modeled with Poissonian noise for photoelectron counts from the star, sky, and dark current, while readout noise (RON) is applied through a  Gaussian random distribution with an rms of the respective camera. RON is quoted as 0.3~e$^-$ with readout time (ROT) 0.039\,s for the ORCA-Quest\,2, and 11.7~e$^-$ with a ROT of 1.65\,s for the iKon-L at 3\,MHz unbinned operation, respectively. Modeling pixel-to-pixel response variation across the sensor was not performed because the resulting uncertainty can be assumed to become negligible after flat-field correction \citep{Hu2025}. As the focus is on the importance of readout noise and sampling, we have furthermore not attempted to simulate the effects of scintillation noise and seeing which should affect the photometry with identical apertures for the two compared detectors in very similar ways.

\begin{figure}[t!]
\centering
\includegraphics[width=\columnwidth]{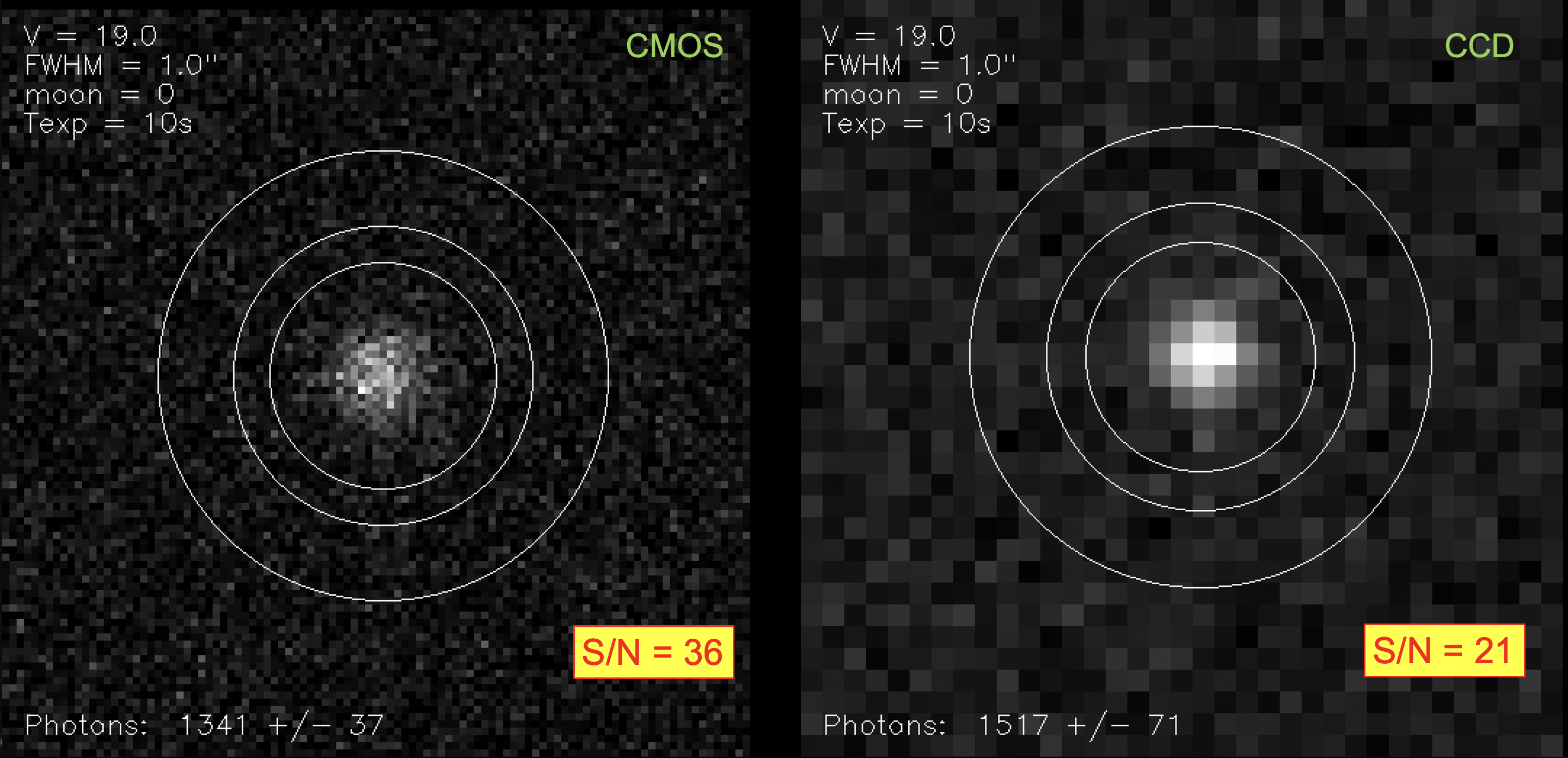}
\includegraphics[width=\columnwidth]{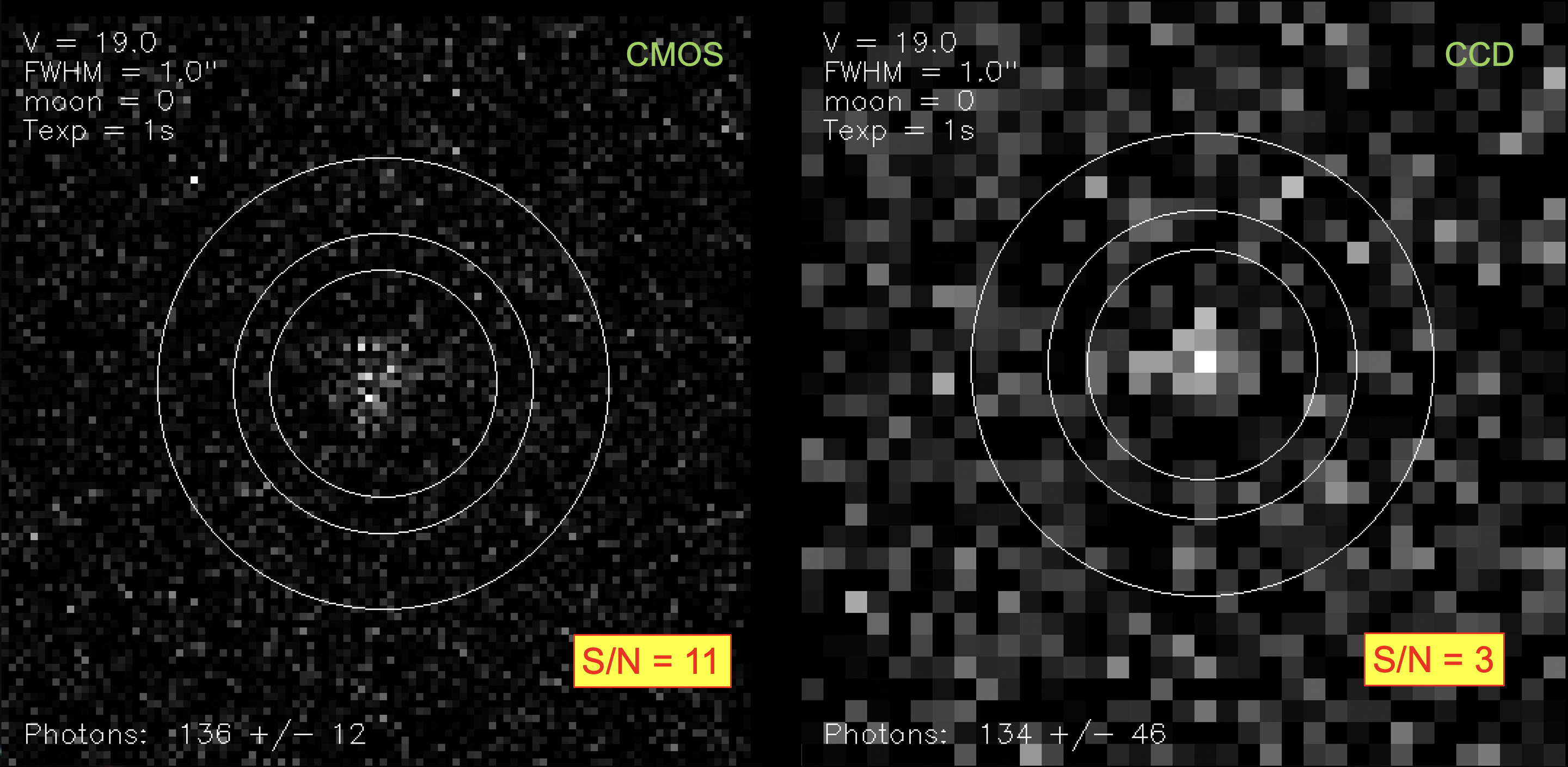}
\includegraphics[width=\columnwidth]{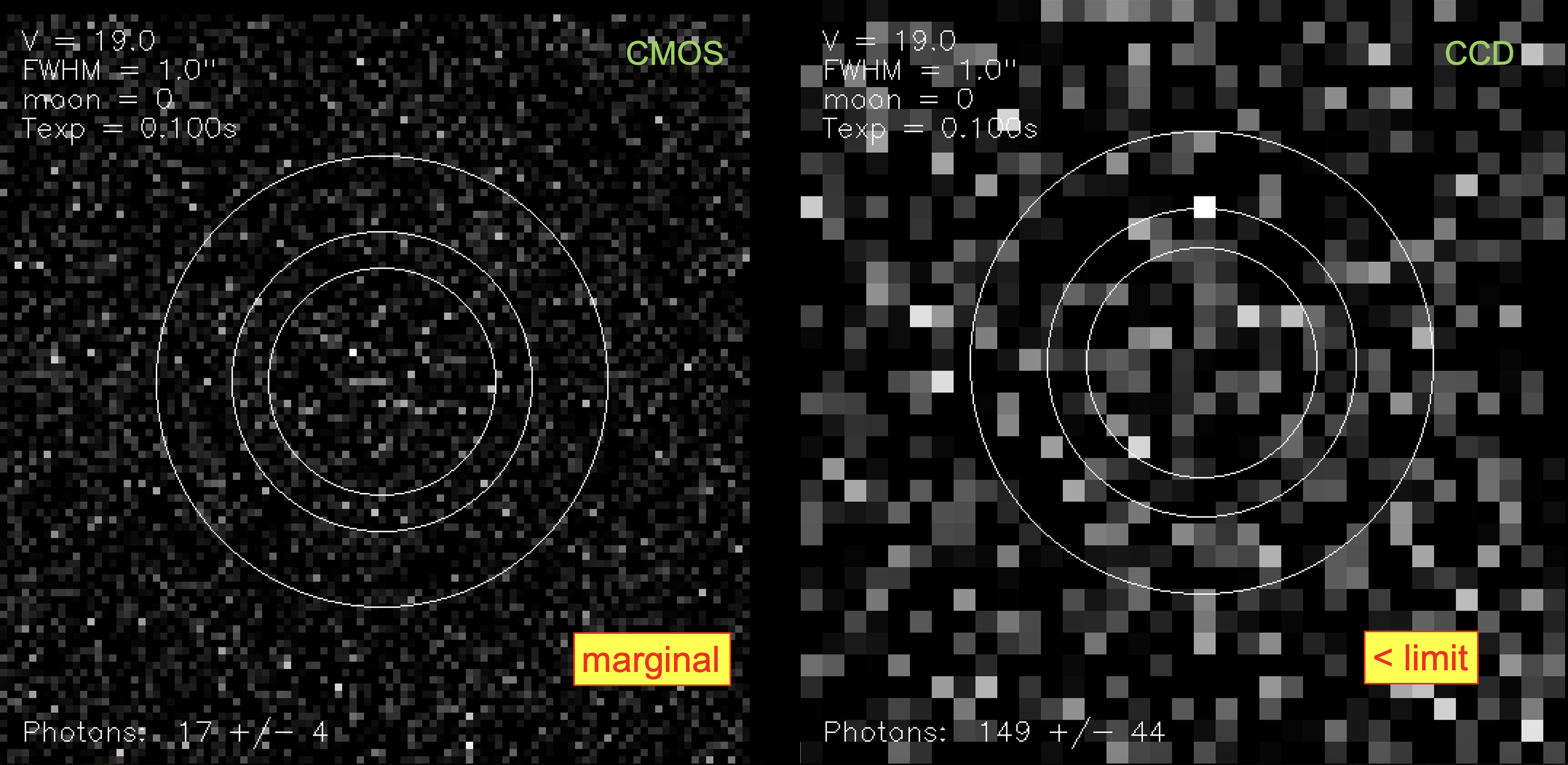}
\caption{Simulated images in the V band for qCMOS (left) and CCD sensor (right), showing a sequence of decreasing exposure times 10\,s, 1.0\,s, and 0.1\,s. The resulting SNR from DAOPHOT photometry on the stellar images is highlighted with yellow labels.}
\label{fig:simul-short}
\end{figure}

Figure~\ref{fig:V18_10s} shows an example for the simulation of a star of magnitude V = 18.0, observed with an exposure time of 10\,s at new moon, with a seeing of 1.0\,arcsec\ FWHM. The different pixel sizes of the qCMOS and the CCD sensor, respectively, are immediately obvious. From images like the ones of this example, the stellar flux and its uncertainty are then measured with DAOPHOT aperture photometry, as indicated by the concentric circles in Fig.~\ref{fig:V18_10s} for the aperture and the sky annulus around the centroid of the star. The number of detected photons and its uncertainty are printed in the lower left corner of the plot.

Although at first glance the more scattered visual appearance of the qCMOS image may intuitively suggest a better result for the CCD, the measured number of detected photons and its uncertainty show the opposite: there is a factor of $\sim$1.4~SNR advantage for the qCMOS over the CCD. 

The example is a good case to illustrate how very low RON wins, even in the event of extreme oversampling: if $N$ is the number of qCMOS pixels that are equivalent to the size of a CCD pixel (in this case approximately $N =  9$), then the RON of the N contributing pixels adds in quadrature to $\sqrt{N} \times$\,RON$_\mathrm{qCMOS}$, i.e.\ $3 \times 0.3$~e$^-$. The resulting binned pixel noise of 0.9~e$^-$ must be compared to the RON of 11.7~e$^-$ at 3~MHz readout for the equivalent CCD pixel, which means a significant gain of 13. 

Even for the slow scan mode of 50~kHz of the CCD with RON = 2.9~e$^-$, there is still a gain of 3. One can go one step further and compare a 2$\times$2 on-chip binned CCD superpixel with 36 equivalent qCMOS pixels, resulting in 1.8~e$^-$ versus 11.7~e$^-$, i.e.\ still a huge gain for the qCMOS.

Figure~\ref{fig:simul-short} illustrates this from another perspective: the same star of magnitude V = 19.0 observed with exposures times of 10~s, 1.0~s, and 0.1~s, respectively. The numbers of detected photons printed in the lower left corner of the images allow to follow the steps of a factor of 10, e.g.\ for the qCMOS: $1341\pm37$, $136\pm12$, and $17\pm4$ photons, which is reasonably in line with the expectation. The last image would suggest to the eye that it is pure noise, however, the DAOPHOT measurement informs otherwise: a marginal SNR = 4 detection. For comparison, the CCD has already reached the detection limit at this short exposure time.

Obviously, the advantage diminishes  with long exposure times needed for very faint objects and background-limited observations. However, the gain for high cadence with short exposure times is clear.

\begin{figure}[t!]
\centering
\includegraphics[width=\columnwidth,bb=20 70  750 500,clip]
{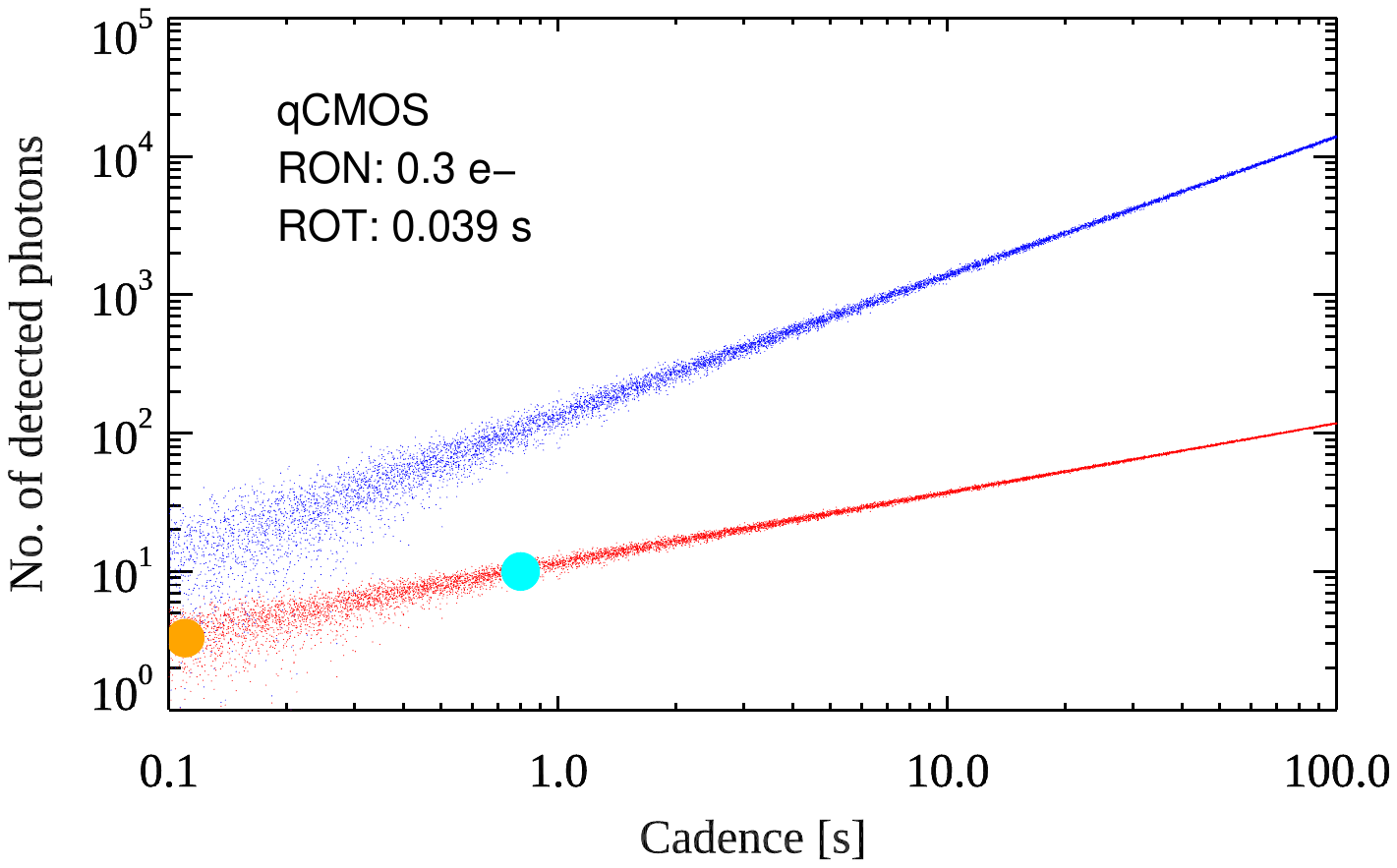}
\includegraphics[width=\columnwidth,bb=20 70  750 500,clip]{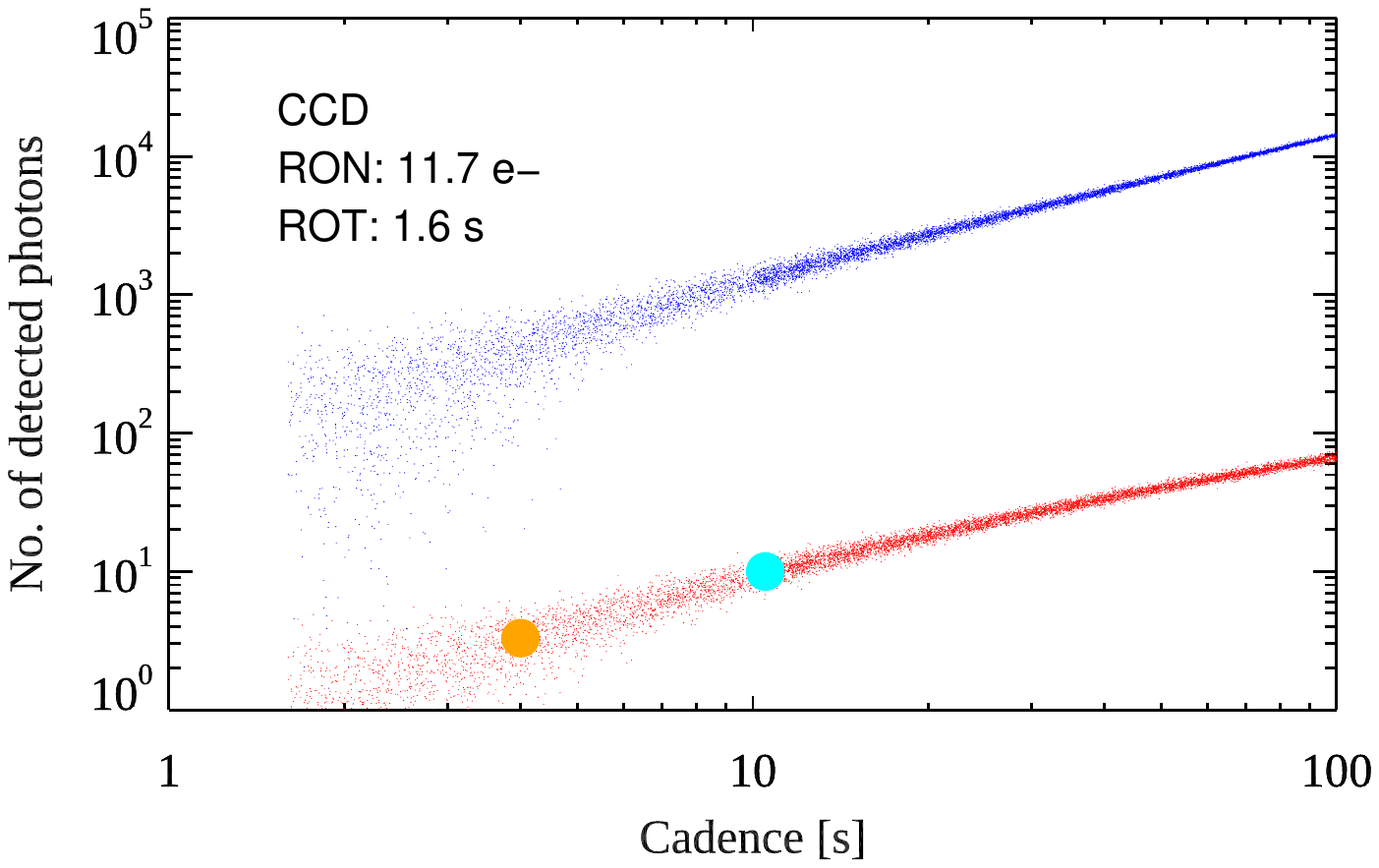}
\caption{10\,000 realizations of simulation of 19th magnitude star under excellent observing conditions, plotted as number of 
    detected photons (blue) versus cadence time step in seconds. Red dots: Corresponding SNR distribution. The circle in orange indicates the detection limit, defined as SNR = 3. The circle in cyan marks the region where exposures reach SNR = 10.}
\label{fig:simul-10K}
\end{figure}

It is important to emphasize the difference between shutter open time for a single exposure and the effective exposure time for a cadence of time series exposures, where the detector ROT must be subtracted from the time step between two exposures, thus clearly reducing the CCD efficiency when the time step is short enough to become comparable to the ROT. We note that for the qCMOS the so-called {\em rolling shutter} effect was not considered in the simulation, but will be investigated as part of a future lab test.

The scatter plots for a total of 10\,000 simulations (Fig.~\ref{fig:simul-10K}) illustrate the situation for time series observations with a cadence between 0.1\,s and 100\,s for the qCMOS (top) and the CCD (bottom), respectively. The observing conditions were chosen to be photometric, new moon, and seeing 1.0\,arcsec\ FWHM. Each blue dot represents the number of photons recovered with DAOPHOT photometry from the simulated star image with a randomly chosen cadence time step, while a corresponding red dot gives the SNR as also reported by DAOPHOT. The range of data points on the abscissa is truncated for the CCD at 1.6\,s because a cadence shorter than the readout time does not make sense, while the range for the qCMOS extends down to a cadence of 0.1\,s. Moreover, the SNR for the qCMOS is everywhere greater than the one for the CCD. The orange circles indicate the detection limit, while the circles in cyan point to the cadence where SNR = 10 is reached: about 10\,s for the CCD and 0.8\,s for the qCMOS. The CCD is readout noise limited in the fast readout regime, while the qCMOS is photon shot noise limited throughout. It is worthwhile noting that the direct one-to-one comparison of the two sensors is an extreme case: it would be even more in favor of the qCMOS if the plate scale was adapted with a focal reducer lens.

\section{Commissioning at Calar Alto} \label{sec:commissioning}

During a period of four nights allocated from June~9 to~12, 2025, the first ORCA-TWIN camera was commissioned at the  1.23m RC telescope of Calar Alto Observatory in southern Spain. The parameters of the telescope, the resulting pixel scale, and the field-of-view are given in Table~\ref{tab:telparam}.

\begin{table}[h!]
\centering
\caption{Telescope parameters}
\begin{tabular}{lc}
\hline
diameter primary mirror       &  1.23~m \rule{0pt}{12pt} \\ 
diameter central obstruction  &  0.582~m  \\ 
focal length                  &  9.808~m  \\
effective collecting area	  &  0.922~m$^2$ \\
focal ratio	                  &  1/8 \\
telescope field-of-view	      &  90.0\,arcmin \\
aberration-free field-of-view &  15\,arcmin \\
plate scale	                  &  20.9\,arcsec mm$^{-1}$ \\
projected pixel size	      &  0.096\,arcsec $\times$ 0.096\,arcsec \\
detector field-of-view	      &  6.6\,arcmin $\times$ 3.7\,arcmin \rule[-3pt]{0pt}{12pt} \\
\hline
\end{tabular}
\label{tab:telparam}
\end{table}

\begin{figure}[h!]
\centering
\includegraphics[width=\columnwidth]{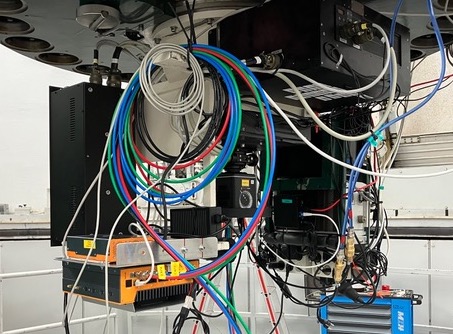}
\caption{ORCA-Quest\,2 camera mounted to Cassegrain focus of CAHA 1.23m telescope. The chiller and the GPS antenna are mounted outside of the scope of the picture.}
\label{fig:cassegrain-mount}
\end{figure}

The ORCA-Quest\,2 camera was mounted at the Cassegrain focus of the 1.23m telescope (Fig.~\ref{fig:cassegrain-mount}) via a C-mount to 2~inch eyepiece adapter and additionally secured with a clamping mechanism to avoid any unwanted movement during telescope operation due to its relatively heavy mass of 3.8~kg. For data acquisition we use the Active Silicon Firebird Quad CXP-6 Frame Grabber installed in an OnLogic Karbon 803 high performance rugged control computer that was mounted at the telescope mechanical interface. The camera is operated in water-cooled mode enabling to reach minimum sensor temperatures of about $-37$~$^\circ$C. As a water recirculating chiller, the Tark Thermal Solutions (formerly Laird Thermal Systems) NRC400-T0-00-PC2 is used, operating at coolant temperature of 20~$^\circ$C and a flow of 0.45~L~min$^{-1}$. For GPS time synchronization the Protempis ThunderBolt GPS Disciplined Clock is employed, whose bullet antenna was placed on the outside railing of the dome.

\section{Observations} \label{sec:results}

\begin{figure}[t!]
\centering
\includegraphics[width=\columnwidth]{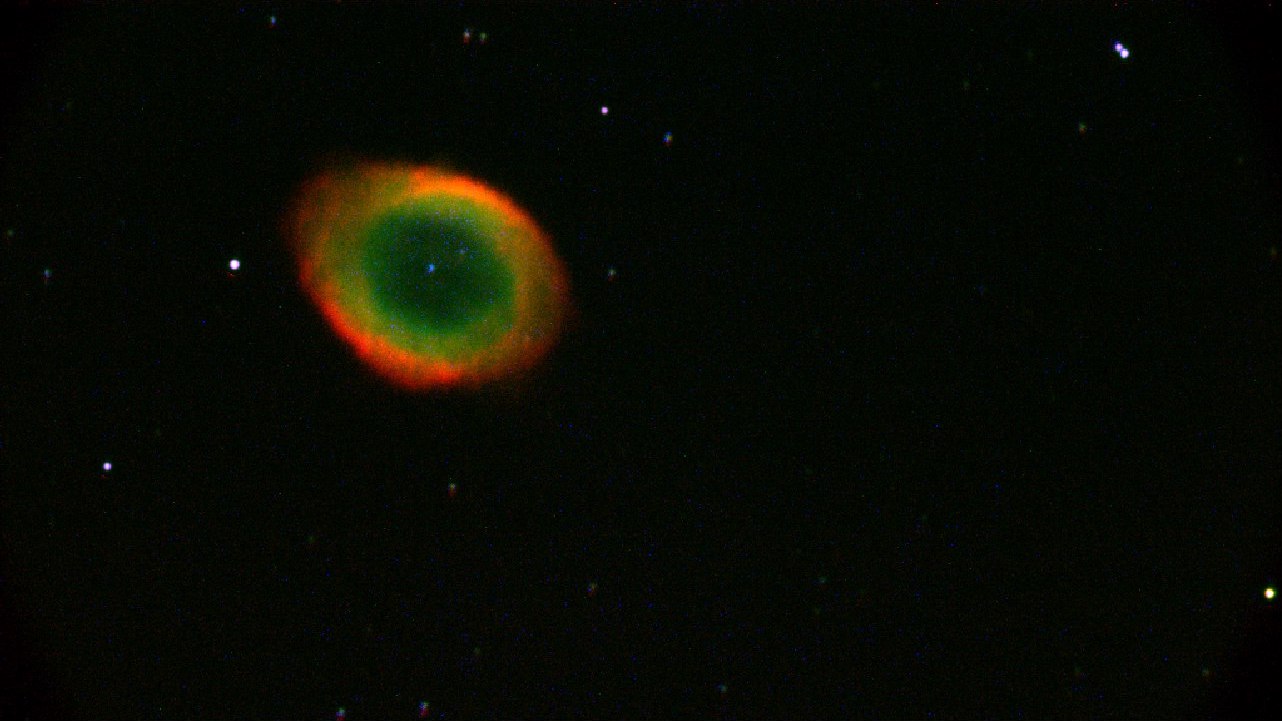}
% ,bb=20 70  750 500, clip
\caption{First-light image: color composite from broad- and narrow-band filter exposures of Ring Nebula M57. Orientation: North up, East left.}
\label{fig:M57}
\end{figure}

The commissioning campaign was scheduled around full moon on June 11, 2025. Weather conditions were unfavorable through the entire run, exacerbated by Calima, a dusty wind from the Sahara desert that prevented observations for most of the time. However, since the camera was swiftly made operational at daytime before the first night and the environmental conditions permitted to open the dome, first light was obtained on June 9, 2025. Figure~\ref{fig:M57} presents a RGB color composite of the Ring Nebula M57 in Lyra, obtained between 22:51~UT and 23:51~UT. Table~\ref{tab:RGB} lists the exposures that have contributed to the image.

\begin{table}[h!]
\centering
\caption{Details of first-light image}
\begin{tabular}{ll}
\hline
\multicolumn{2}{l}{Filters}    \rule[-4pt]{0pt}{15pt}     \\ 
\hline
R: &	[S\,II] $\lambda$6717/$\lambda$ 6731 narrow-band filter \rule{0pt}{12pt} \\
G: &	SDSS g broad-band filter: [O\,III] $\lambda$ 5007/$\lambda$ 4959, H$\beta$ \\
B: &	He\,II $\lambda$ 4686 narrow-band filter 
\rule[-4pt]{0pt}{10pt}\\
\hline
\multicolumn{2}{l}{Exposure times}  \rule[-4pt]{0pt}{15pt}  \\
\hline
R:	& $5 \times 180$~s \rule{0pt}{12pt} \\
G:	& $10 \times 10$~s  \\
B:	& $30 \times 10$~s \rule[-4pt]{0pt}{10pt} \\
\hline
\end{tabular}
\label{tab:RGB}
\end{table}

In the absence of an autoguider, the drift between exposures was compensated by registering the R and B images to the G frames. Moreover, the six brightest stars were processed with aperture photometry and registered with an equalized PSF to the G frame centroids to compensate for subtle wavelength dependent seeing differences, tracking effects, and aberrations.

\begin{figure}[h!]
\centering
\includegraphics[width=\columnwidth,
bb=100 40  1350 730, clip]{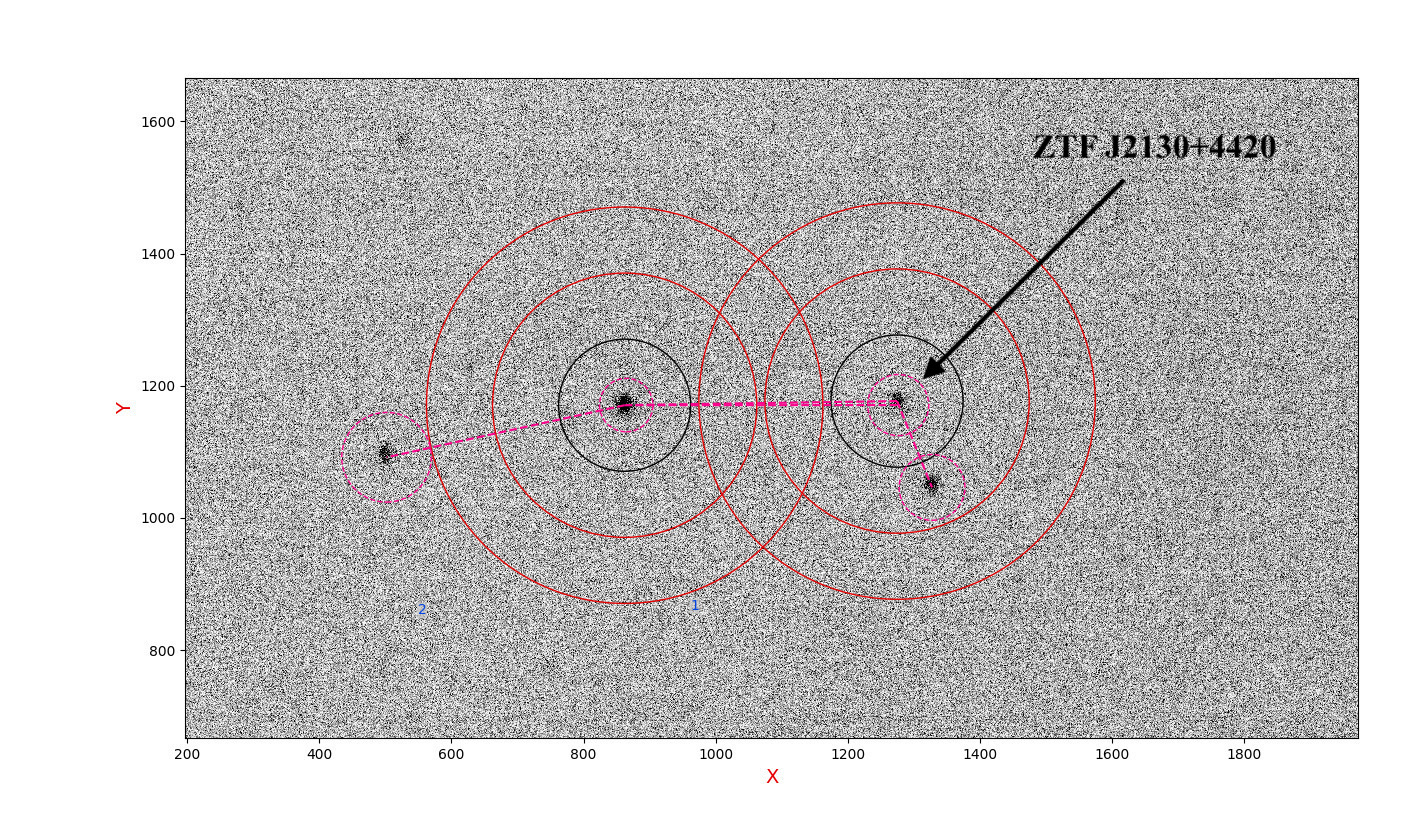}
\caption{Direct image of ZTF J2130+4420 with comparison star. The circles indicate the star apertures (black) and sky annuli (red) for aperture photometry. Orientation: North up, East left.}
\label{fig:ZTF-map}
\end{figure}

\begin{figure}[t!]
\centering
\includegraphics[width=\columnwidth]{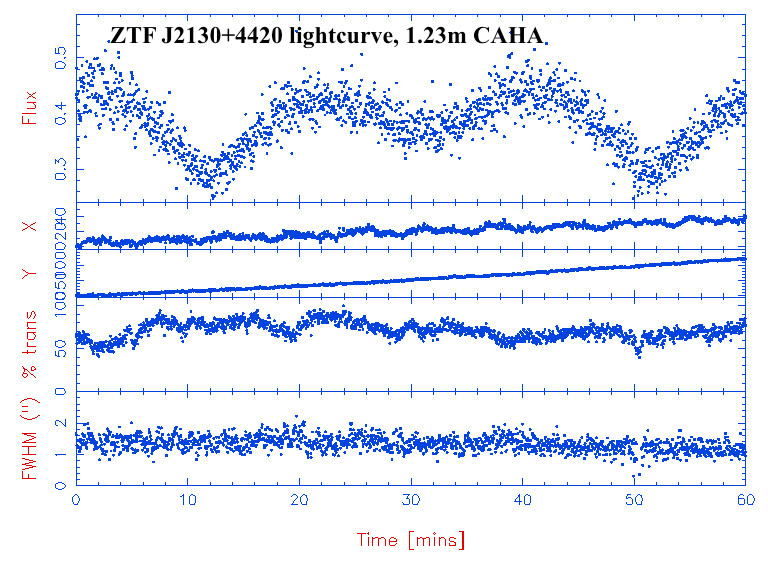}
\caption{Light curve for ZTF J2130+4420 obtained during the observing run with 
    the HiPERCAM data reduction software (top panel, a 60 minute fraction of the observed period is shown). Second and third panels: telescope drift in $X$ and $Y$ coordinates shown in Fig.~\ref{fig:ZTF-map}. Fourth panel: variation of transparency of the atmosphere. Bottom: seeing FWHM in arcseconds.}
\label{fig:ZTF-lightcurve}
\end{figure}

In line with the scientific objective of the ORCA-TWIN project, a primary goal of the first observations was to test the capability for high cadence time resolved photometry. To this end, several short period binaries were selected, out of which the ultra-compact Roche-lobe filling hot subdwarf binary ZTF\,J2130+4420 \citep{Kupfer2020a} turned out to be observable on June 11, 2025, albeit under harsh conditions: full moon, scattered clouds, Calima. The object with a mean magnitude of $g\approx15.45$ could be observed in the Gunn\,g band from 01:05~UTC with a cadence of 2~s, readout time of 0.01~s at 4$\times$4-pixel binning, and a total of 3047 exposures, to cover more than two periods of $P = 39.34$ ~min. Data reduction was performed during the night on-the-fly by AJB using the HiPERCAM pipeline.\footnote{\url{https://github.com/HiPERCAM}} A direct image of the object and a comparison star for differential aperture photometry is shown in Fig.~\ref{fig:ZTF-map}. A fraction of the light curve is plotted in Fig.~\ref{fig:ZTF-lightcurve}. The atmospheric transmission curve illustrates the presence of clouds and very non-photometric conditions. Despite the adverse conditions, differential photometry has enabled to compensate for the variable extinction and derive a meaningful photometry. The result compares well with the discovery light curve, Fig.~1 in \citet{Kupfer2020a}.

The object is interesting as a potential gravitational wave source and a possible candidate for a future supernova type Ia explosion. A forthcoming paper presents the full analysis of the light curve obtained in a follow-up observing run from September 2025 under good observing conditions \citep{Teckenburg2025}.

\begin{figure}[t!]
\centering
\includegraphics[width=\columnwidth]{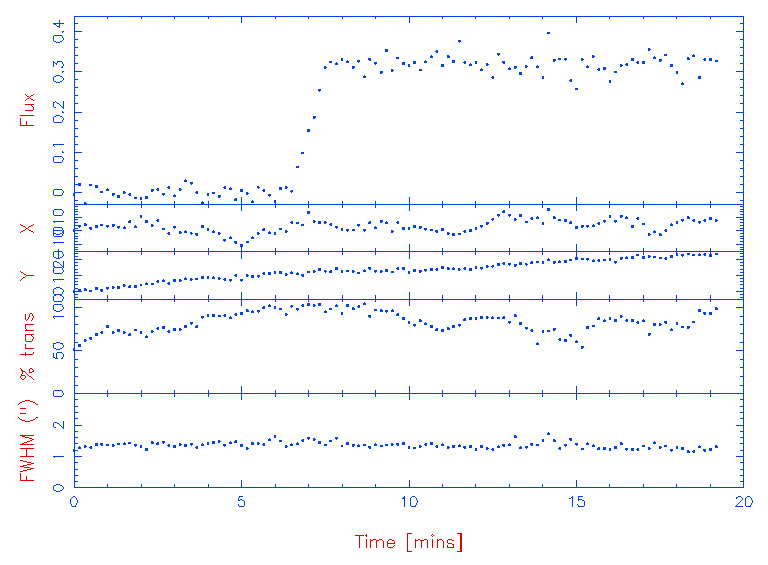}
\caption{Light curve for NN Ser. Due to a temporary technical problem related to the dome mechanism, the eclipse was only partially covered, and the ingress phase missed. For explanation see Fig.~\ref{fig:ZTF-lightcurve}.}
\label{fig:NNSer-lightcurve}
\end{figure}

As another test case the short-period eclipsing pre-cataclysmic binary system NN\,Ser  ($P = 3.12$~h) could be observed partially during eclipse on June 09, 2025. NN\,Ser consists of a hot WD (0.535~$M_\sun$) and a main-sequence dwarf of spectral-type M4 (0.111~$M_\sun$) with a very deep eclipse \citep{Haefner1989}. \citet{Oezdoenmez2023} present modern CCD observations of the system and state that in eclipse the brightness decreases by 5.8~mag from V $\approx$ 17~mag. An analysis of the O--C diagram since 1988 reveals an orbital period variation that the authors explain with the presence of a planet of at least 9.5 Jupiter masses. ORCA-Quest\,2 observations were obtained from 22:07~UT as a series of 116 frames with an exposure time of 10~s using the Gunn\,g filter. Unfortunately, the ingress into eclipse was missed because of a delay caused by a temporary failure of the telescope dome mechanism. 

As for ZTF\,J2130+4420, preliminary data reduction was performed immediately during the night. The light curve shown in Fig.~\ref{fig:NNSer-lightcurve} is in good agreement with Figure~1 in \citet{Oezdoenmez2023}. Again, the non-photometric conditions can be appreciated from the variation of atmospheric transmission on the order of 50\%. The case for high time resolutions is particularly important for this system, as the O--C diagram analysis requires high precision, i.e.\ sufficiently many data points on the steep ingress and egress phases.

\section{Conclusions} \label{sec:conclusions}

Recent years have seen increasing interest in photon counting image sensors, both in the visual, and in the near infrared, for ground based astronomy and applications in space. The commercial availability of the turnkey-ready qCMOS camera system ORCA-Quest\,2 from Hamamatsu has allowed the emerging German Center for Astrophysics to engage in this promising technology early on, and to launch the ORCA-TWIN pilot project, while creating a new center for technology, and building first collaborations with the community in Hamburg, Potsdam, Garching, Granada, and Rome. With support from Calar Alto Observatory, a fast track installation of the first ORCA-TWIN camera at the 1.23m telescope has been accomplished at a four-night commissioning run in June 2025, three months after kick-off. Despite poor, non-photometric observing conditions near full moon, first light was obtained successfully, and the feasibility of high-cadence photometry was demonstrated, in agreement with the predictions from numerical simulations. 

The main result is the confirmation that despite the small pixel size and significant oversampling of the stellar PSF, albeit a relatively small FoV for the given sensor format of $4096\times2304$ pixels, the qCMOS sensor is a competitive device at 1m-class telescopes. Moreover, the small noise penalty for post-readout binning opens interesting avenues for scene-dependent image processing, and even applications in spectroscopy. For future developments of this technology, a larger pixel pitch, e.g. 15 $\mu$m, would be desirable. Such a value is typical for CCDs and would translate to a projected pixel size of ~ 0.3\,arcsec at the 1.23m Telescope, i.e. Nyquist sampling for sub-arcsec seeing.

In the meantime, two more science observing runs were conducted at CAHA in September and November 2025 whose results will be published in the near future. 

Forthcoming Papers~II and~III will report results from characterization of the ORCA-Quest\,2 camera in the DZA detector laboratory and performance during the first science observing runs, and from commissioning at the Teide Observatory as well as first results from simultaneous observations using the twin configuration, respectively.

\section*{Acknowledgments}

%We thank the anonymous referee for the useful suggestions. 
Based on data obtained from the Calar Alto Observatory.
MMR, PR, SV, and MK acknowledge financial support from BMFTR under grant 03WSP1745 and excellent administrative support from Projekt\-träger Jülich. HD was supported by the Deutsche Forschungsgemeinschaft (DFG) through grants GE2506/17-1 and GE2506/9-2. We acknowledge technical support from Calar Alto Observatory and Leibniz Institute for Astrophysics Potsdam (AIP). J.\,L.\,O. acknowledges support from the Spanish projects PID2020-112789GB-I00 (AEI), Proyecto de Excelencia de la Junta de Andalucía PY20-01309 and Severo Ochoa grant CEX2021-001131-S (MCIN/AEI/10.13039/501100011033).
This research was supported by Deutsche Forschungsgemeinschaft (DFG, German Research Foundation) under Germany’s Excellence Strategy -- EXC 2121 ``Quantum Universe'' -- 390833306 and co-funded by the European Union (ERC, CompactBINARIES, 101078773). Views and opinions expressed are, however, those of the author(s) only and do not necessarily reflect those of the European Union or the ERC. Neither the European Union nor the granting authority can be held responsible for them.

%\subsection*{Author contributions}
%This is an author contribution text. This is an author contribution text. This is an author contribution text.  

\subsection*{Financial disclosure}

None reported.

\subsection*{Conflict of interest}

The authors declare no potential conflict of interests.

\section*{Supporting information}

The data that support the findings of this study are openly available at \url{https://cloud.aip.de/index.php/s/BTzNTJZ4amQGPyW}.

\clearpage

\appendix

\section{Numerical Simulations of qCMOS versus CCD performance}
\label{appendix:qcmos-simulations}

The simulation of stellar images on the qCMOS and CCD cameras is coded in IDL scripts, creating images with the plate scale of a 1.23m, $f/8$ focal ratio telescope, and a selectable pixel size. The resulting arrays are restricted to windows of a projected size of approximately 10\,arcsec $\times$ 10\,arcsec. The exact values of the parameters are listed in Table~\ref{tab:simparam}.

On these arrays, the surface brightness distribution of a star, centered on the array, is realized with a Gaussian PSF with selectable seeing FWHM, the integral flux being normalized to unity. A more accurate Moffat profile is not deemed necessary for the simple aperture photometry used in the simulation. The physical star and sky photon counts per pixel are computed for a given magnitude in the desired broadband filter by multiplying the array with the total photon flux above the atmosphere tabulated in Table~\ref{tab:photonflux} , using the relation 1~Jy $= 1.51 \times 10^7$ photons~m$^{-2}$~s$^{-1}$\,$\Delta\lambda/\lambda$, multiplied with the exposure time and the throughput of the system, which is given by the QE of the detector, atmospheric extinction, the effective light-collecting area of the telescope, and reflection losses of optical surfaces. The sky background is added using the surface brightness estimates for different Moon phases as listed in Table~\ref{tab:photonflux}, as well as the dark current for the selected exposure time. 

The resulting number of detected photons per pixel is then made subject to a Poissonian and Gaussian noise model, the former applicable to detected photons and dark current, the latter for the readout noise of the camera. 

In order to create an empirical result for the detected signal and its uncertainty, aperture photometry with DAOPHOT \citep{Stetson1987} is applied to the final image. The aperture and inner and outer sky annulus radii are given in Table~\ref{tab:simparam}.

% #########################################################

\begin{table}[t!]

\caption{Simulation parameters \label{tab:simparam}}
\begin{tabular}{|ll|}
\hline
\multicolumn{2}{|l|}{\bf Telescope~~~~~~~~~~~~~~~~~~~\rule[-4pt]{0pt}{15pt}}  \\
\hline
diameter primary mirror       &  1.23~m\rule{0pt}{12pt}   \\ 
diameter central obstruction  &  0.582~m  \\ 
focal length                  &  9.808~m  \\
plate scale                   & 20.9 arcsec~mm$^{-1}$\rule[-4pt]{0pt}{10pt} \\
\hline\hline
\multicolumn{2}{|l|}{\bf qCMOS\rule[-4pt]{0pt}{15pt}}       \\ 
\hline
pixel size           &  $4.6~\mu$m $\times$ 4.6~$\mu$m\rule{0pt}{12pt}  \\
projected pixel size on the sky &  $0.096\times0.096$ arcsec$^2$ \\
readout noise        &  0.3~e$^-$            \\      
readout time         &  0.039~s              \\
dark current ($-20$~$^\circ$C) & 0.016~e$^-$~s$^{-1}$~pixel$^{-1}$ \\
window               & $103\times103$ pixels \\
                     & $9.96 \times 9.96$~arcsec$^2$ \\
DAOPHOT APR          & 15.51~pixels \\
                     & 1.5 arcsec \\
DAOPHOT SKYRAD       & [20.67, 31.01] pixels \\                                       & [2.0 arcsec, 3.0 arcsec]\rule[-4pt]{0pt}{10pt} \\   
\hline\hline
\multicolumn{2}{|l|}{\bf CCD\rule[-4pt]{0pt}{15pt}}         \\ 
\hline
pixel size         &  $13.5~\mu$m $\times$ 13.5~$\mu$m\rule{0pt}{12pt}  \\
projected pixel size on the sky &  $0.282 \times 0.282$ arcsec$^2$ \\
readout noise      &  11.7~e$^-$            \\  
readout time       &  1.6~s at 3~MHz                \\
dark current ($-80$~$^\circ$C) & 0.00013~e$^-$ s$^{-1}$ pixel$^{-1}$ \\
window               & $35\times35$ pixels \\
                     & $9.94 \times 9.94~$arcsec$^2$ \\
DAOPHOT APR          & 5.28~pixels \\
                     & 1.5 arcsec \\
DAOPHOT SKYRAD       & [7.04, 10.57] pixels \\                                       & [2.0 arcsec, 3.0 arcsec]\rule[-4pt]{0pt}{10pt} \\ 
\hline
\end{tabular}~~~~~~~~~~~~~~~~~~~~~~~~~~~~~~~~
\end{table}

% #########################################################
\begin{table}[t!]
\caption{Photon flux for different filters \label{tab:photonflux}}
\begin{tabular}{|lllcc|}
\hline
\multicolumn{5}{|l|}{\bf Photon flux [photons~s$^{-1}$ ~m$^{-2}$ above atmosphere]\rule[-4pt]{0pt}{15pt}}  \\
\hline
Band & $\lambda$ & $\Delta\lambda/\lambda$ & $F(m=0)$ [Jy] & References\rule[-4pt]{0pt}{15pt} \\
\hline
U  &  0.36  &  0.15  &  1810  &  (1)\rule{0pt}{12pt} \\
B  &  0.44  &  0.22  &  4260  &  (1) \\
V  &  0.55  &  0.16  &  3640  &  (1) \\
R  &  0.64  &  0.23  &  3080  &  (1) \\
I  &  0.79  &  0.19  &  2550  &  (1) \\
J  &  1.26  &  0.16  &  1600  &  (2) \\
H  &  1.60  &  0.23  &  1080  &  (2) \\
K  &  2.22  &  0.23  &   670  &  (2) \\
g  &  0.52  &  0.14  &  3730  &  (3) \\
r  &  0.67  &  0.14  &  4490  &  (3) \\
i  &  0.79  &  0.16  &  4760  &  (3) \\
z  &  0.91  &  0.13  &  4810  &  (3)\rule[-4pt]{0pt}{10pt} \\
\hline
\multicolumn{5}{l}{~} \\
\multicolumn{5}{l}{Ref. (1): \citet{Bessel1979}} \\
\multicolumn{5}{l}{Ref. (2): \citet{Campins1985}} \\
\multicolumn{5}{l}{Ref. (3): \citet{Schneider1983}} \\
\multicolumn{5}{l}{~} \\
\end{tabular}

\vspace{5mm}
\begin{tabular}{|cccccc|}
\hline
\multicolumn{6}{|l|}{\bf Sky brightness [mag~arcsec$^{-2}$] with Moon phase\rule[-4pt]{0pt}{15pt}} \\
\hline
\hline
~Phase [d] & U &  B    & V    & R     & I\rule[-4pt]{0pt}{15pt}      \\
\hline
 0  & ~22.0~~ &	~22.7~~  & ~21.8~~  & ~20.9~~ & ~19.9~\rule{0pt}{12pt}  \\
 3  & 21.5 &	22.4  & 21.7  & 20.8 & 19.9   \\
 7  & 19.9 &	21.6  & 21.4  & 20.6 & 19.7   \\
 10 & 18.5 &	20.7  & 20.7  & 20.3 & 19.5   \\
 14 & 17.0 &	19.5  & 20.0  & 19.9 & 19.2\rule[-4pt]{0pt}{10pt}   \\
\hline
\end{tabular}
\end{table}

\begin{figure}[!h]
	\centering
	\includegraphics[width=1.0\columnwidth]{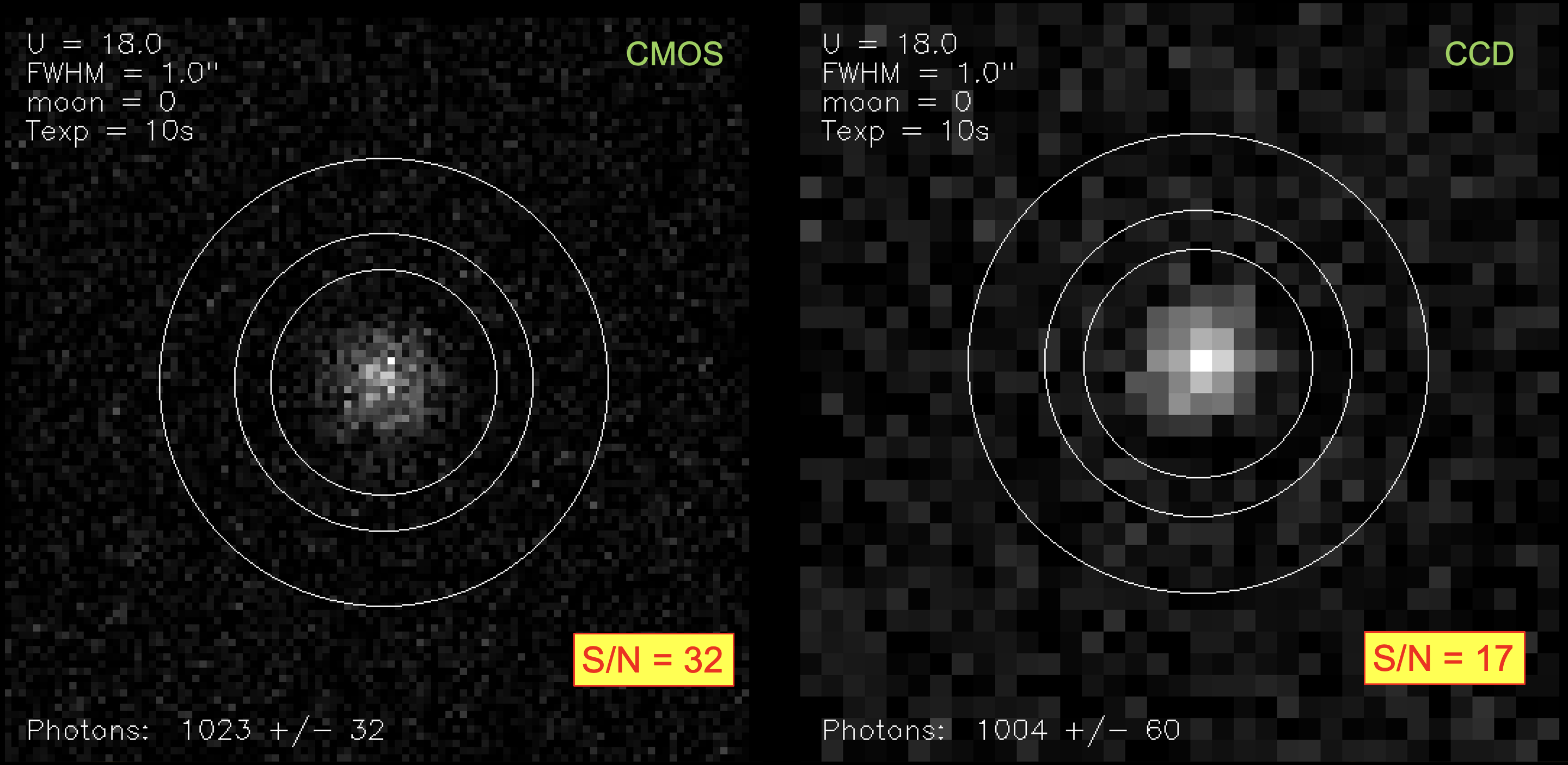}
	\includegraphics[width=1.0\columnwidth]{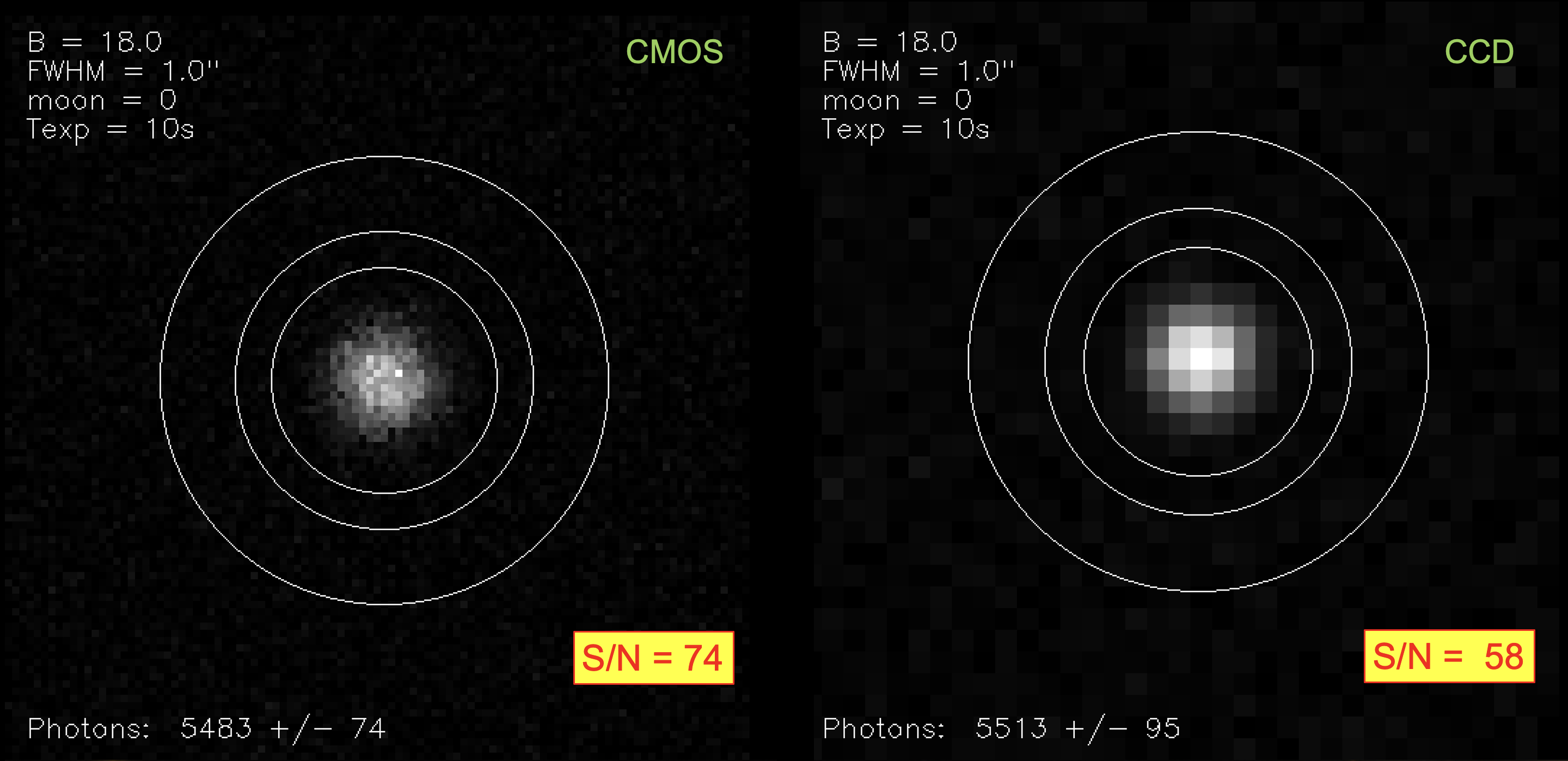}
	\includegraphics[width=1.0\columnwidth]{V18.png}
	\includegraphics[width=1.0\columnwidth]{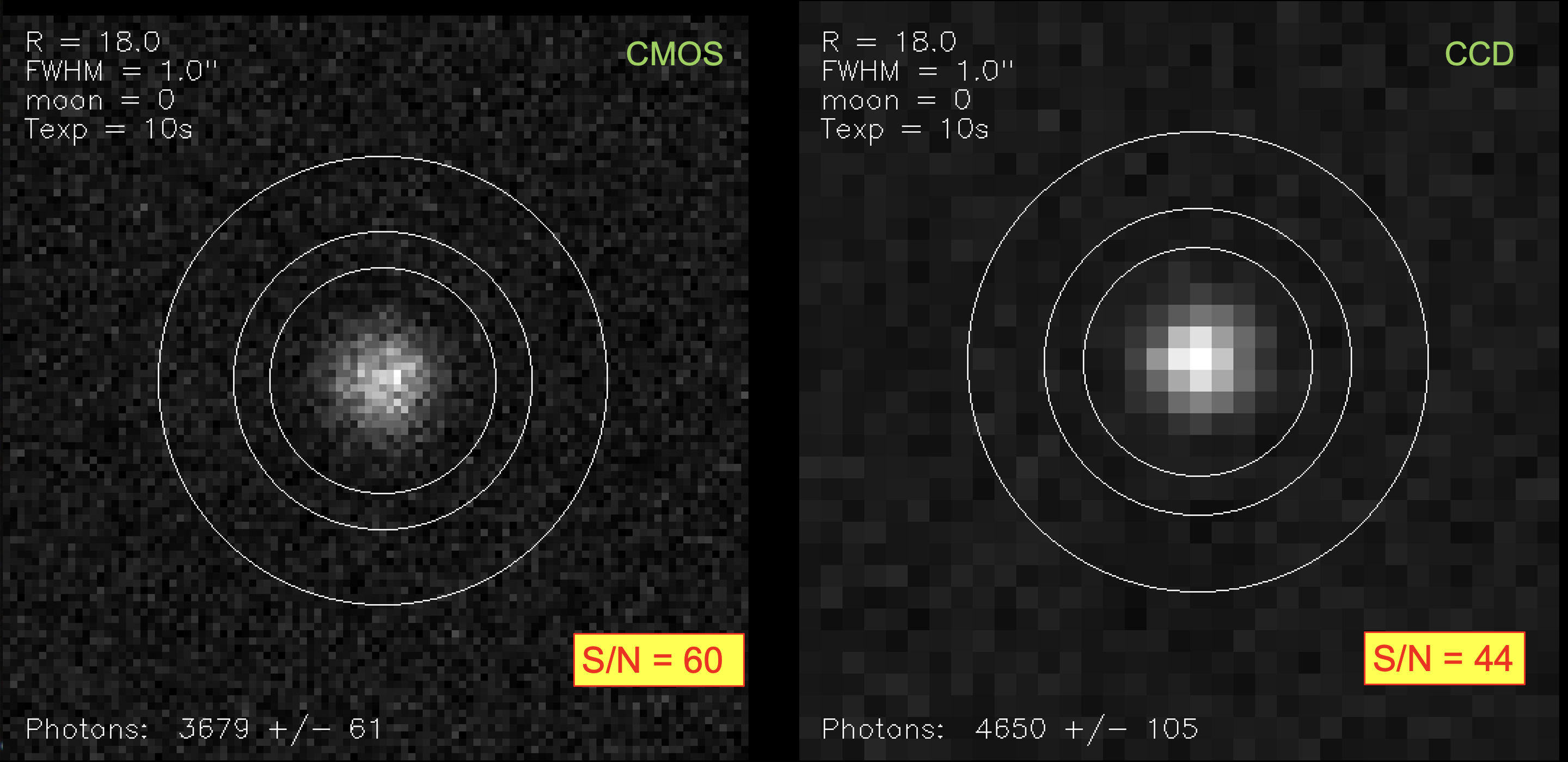}
	\includegraphics[width=1.0\columnwidth]{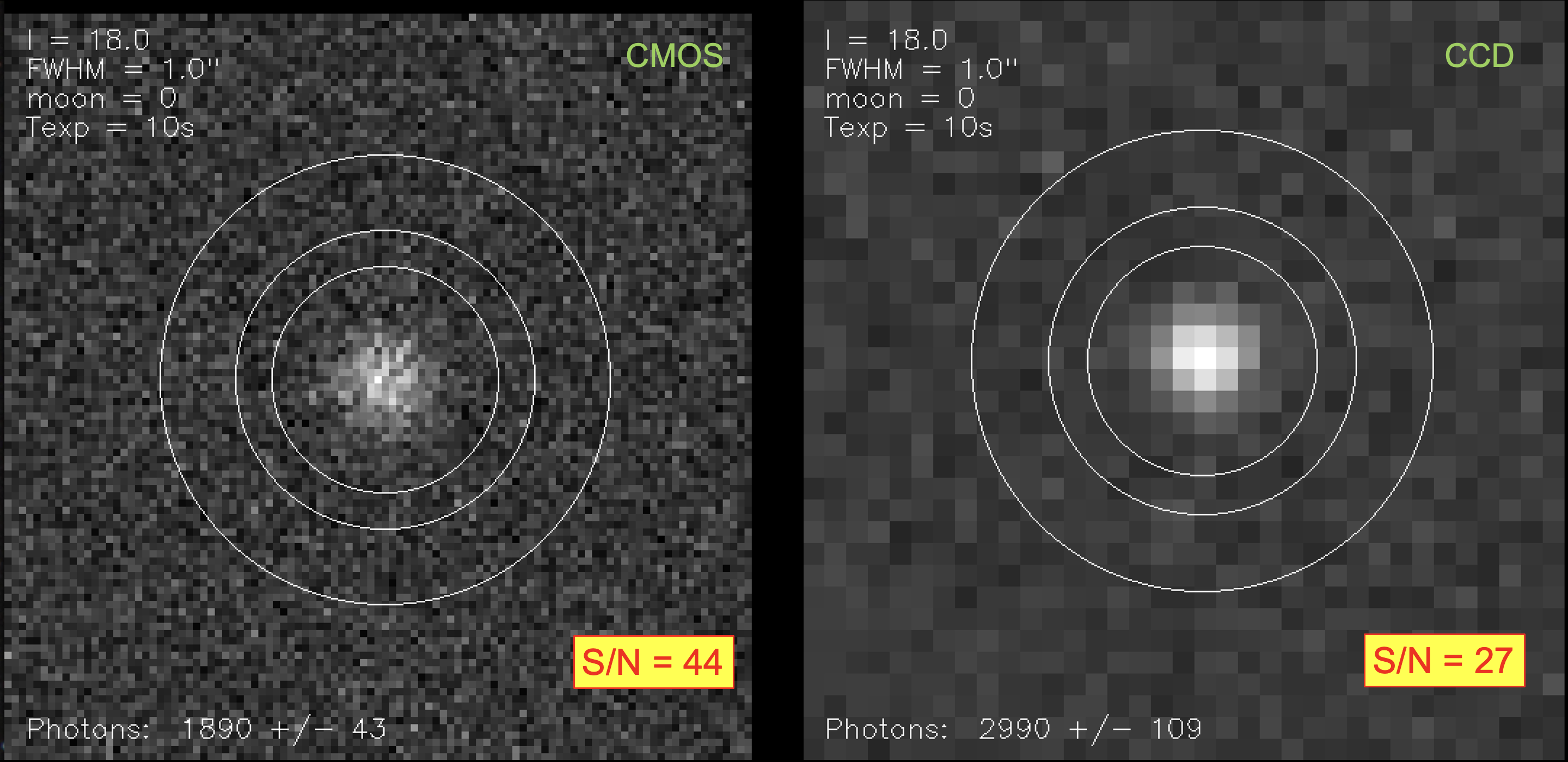}
	\caption{Simulated images for 18th magnitude star at qCMOS (left) and CCD sensor (right), for broadband filters Bessel U, B, V R, I, an exposure time of 10\,s, 1.0~arcsec seeing, and new moon.}
	\label{fig:simul-UBVRI}
\end{figure}

\clearpage

% ########################################################
\nocite{*}% Show all bib entries - both cited and uncited; comment this line to view only cited bib entries;
\bibliography{Wiley-ASNA}%
% ########################################################

%\section*{Author Biography}
%(if applicable)

%\begin{biography}{\includegraphics[width=60pt,height=70pt,draft]{empty}}{\textbf{Author Name.} This is sample author biography text this is sample author biography text this is sample author biography text this is sample author biography text this is sample author biography text this is sample author biography text this is sample author biography text this is sample author biography text this is sample author biography text .}
%\end{biography}

\end{document}